\documentclass[12pt]{amsart}
\usepackage{amsmath,amsthm,amsfonts,amssymb,eucal, hyperref}

\input header.sty
\usepackage{graphicx,psfrag,amssymb,amsmath}

\newcommand{\BZ}{\mathbf Z}

\begin{document}

\hoffset -4pc

\title[Density of states]
{Asymptotic expansion of the integrated density of states of a two-dimensional periodic Schr\"odinger operator}
\author[L. Parnovski \& R. Shterenberg]
{Leonid Parnovski \& Roman Shterenberg}
\address{Department of Mathematics\\ University College London\\
Gower Street\\ London\\ WC1E 6BT UK}
\email{Leonid@math.ucl.ac.uk}
\address{Department of Mathematics\\ University of Alabama at Birmingham\\ 1300 University Blvd.\\
Birmingham AL 35294\\ USA}
\email{shterenb@math.uab.edu}

\date{\today}
\begin{abstract}
We prove the complete asymptotic expansion of the integrated density of states of a
two-dimensional Schr\"odinger operator 
with a smooth periodic potential.
\end{abstract}

\maketitle
\vskip 0.5cm

\section{Introduction}

Let $H$ be a Schr\"odinger operator
\bee\label{operator1}
H=-\Delta+V
\ene
acting in $\R^d$. The potential $V$ is assumed to be infinitely smooth and periodic with
$\Gamma\subset\R^d$ being its lattice of periods. We denote by $\CO=\R^2/\Gamma$ the fundamental quotient of $\Ga$
and by $v$ the $L^\infty$-norm of $V$.
We also denote by $\Gamma^{\dagger}$ a dual lattice to $\Gamma$ and put $\CO^{\dagger}=\R^2/\Gamma^{\dagger}$.
Denote by $\tilde N(\lambda)$ the (integrated)
density of states of the operator $H$. The density of states
is defined by the formula
\bee
\tilde N(\lambda) = \lim_{L\to\infty}
\frac{N(\lambda;H^{(L)}_D)}
{L^d}.
\ene
Here, $H^{(L)}_D$ is the restriction of $H$ to the cube $[0,L]^d$ with the Dirichlet boundary
conditions, and $N(\lambda;A)$ is the counting function of the discrete spectrum of (a bounded below operator with compact resolvent) $A$.
If we denote by $\tilde N_0(\lambda)$ the density of states of the unperturbed operator
$H_0=-\Delta$, one can easily see that for positive $\la$ one has
\bee
\tilde N_0(\lambda)=\frac{1}{(2\pi)^d}w_d\la^{d/2},
\ene
where
\bee
w_d=\frac{\pi^{d/2}}{\Gamma(1+d/2)}
\ene
is the volume of the unit ball in $\R^d$. There is a long-standing conjecture that for large $\lambda$
the density of states of the perturbed operator enjoys the following asymptotic behaviour as $\la\to\infty$:
\bee\label{eq:intr0}
\tilde N(\lambda)\sim \tilde N_0(\lambda)\Bigl(1+\sum_{j=1}^\infty e_j\lambda^{-j}\Bigr),
\ene
meaning that for each $K\in\N$ one has
\bee\label{eq:intr1}
\tilde N(\lambda)=\tilde N_0(\lambda)\Bigl(1+\sum_{j=1}^K e_j\lambda^{-j}\Bigr)+R_K(\lambda)
\ene
with $R_K(\lambda)=o(\la^{\frac{d}{2}-K})$. In these formulas, $e_j$ are real numbers which depend on the potential $V$.
They can be calculated relatively easily using the heat kernel invariants (computed in \cite{HitPol}); they are equal to certain
integrals of the potential $V$ and its derivatives.
Indeed, in the paper \cite{KorPush}, all these coefficients were computed; in particular, it turned out
that if $d$ is even, then $e_j$ vanish whenever $j>d/2$.

So far, formula \eqref{eq:intr0} has been proved only in the case $d=1$ in the paper \cite{ShuSche}.
In the multidimensional case, only partial results are known, see \cite{HelMoh}, \cite{Kar},
\cite{Kar1}, \cite{ParSob}, \cite{Skr}, \cite{Sob}. In particular, in \cite{Sob} it was shown
that when $d=2$ formula \eqref{eq:intr1} is valid with $K=2$ and $R(\lambda)=O(\lambda^{-\frac{6}{5}+\epsilon})$ for any
positive $\epsilon$; in \cite{Kar} it was shown that when $d\ge 3$ formula \eqref{eq:intr1} is valid with $K=1$ and $R(\la)=O(\lambda^{-\de})$ with some small $\de$ when $d=3$ and $R(\la)=O(\lambda^{\frac{d-3}{2}}\ln\la)$ when $d>3$.

The aim of this paper is to establish the complete asymptotic formula \eqref{eq:intr0} in the $2$-dimensional case. Namely,
we will prove that if $d=2$, we have:
\bee\label{eq:intr2}
\tilde N(\lambda)=\frac{1}{4\pi}(\la-b)+O(\la^{-K})
\ene
for each $K\in\N$ as $\la\to\infty$ with
\bee\label{eq:b}
b:=\frac{1}{\vol(\CO)}\int_{\CO}V(x)dx.
\ene
Note that in view of \cite{KorPush}, it is enough to establish that \eqref{eq:intr1} holds for each $K$ with some constants $e_j$; then \eqref{eq:intr2} will follow automatically. Moreover, suppose that we have proved the following asymptotic formula:
\bee\label{eq:intrnew1}
\tilde N(\lambda)=\tilde N_0(\lambda)\Bigl(1+\sum_{j=1}^{2K} e_j\lambda^{-j/2}+\sum_{j=1}^{2K} \hat e_j\lambda^{-j/2}\ln(\la)\Bigr)+o(\la^{\frac{d}{2}-K}).
\ene
Then, applying the same arguments as in \cite{KorPush}, together with some straightforward calculations (one needs to compute the Laplace
transform of $\la^{\alpha}\ln\la$), it is easy to show that \eqref{eq:intrnew1} still implies \eqref{eq:intr2}. Therefore, our aim will be to prove
\eqref{eq:intrnew1}. It was quite surprising for us when we were performing the calculations that the terms containing logarithms were actually
`present' in the asymptotics of $\tilde N$, although the coefficients $\hat e_j$ in front of these terms turned out to be zero.
\ber
The coefficients $\hat e_j$ in front of logarithmic terms can be non-zero if one allows non-local pseudo-differential perturbations $V$. For example, suppose, $\Gamma=\Z^2$ and $V$ is a pseudo-differential operator of order zero with the following symbol:
\bee
v(\bx,\bxi)=[\cos(2\pi x_1)+\cos(2\pi x_2)+\cos(2\pi(x_1-x_2))]\chi_1(|\bxi|)\chi_2(\arg(\bxi)),
\ene
where $\chi_1$ is a smooth cut-off to the interval $[1,+\infty)$, and $\chi_2$ is a smooth cut-off to $[-0.1,\pi/4]$. Then the formula \eqref{eq:intrnew1}
is still valid, with $\hat e_4\ne 0$. This can be seen by repeating the arguments of our paper for non-local operators and a careful computation of all coefficients. Since in our paper we do not consider non-local perturbations, we will not go into more details, but we may return to this example in a further publication.
\enr

The method we apply to establish \eqref{eq:intrnew1} consists of two parts. The first part is, essentially, the method used in \cite{Par} in order to prove the Bethe-Sommerfeld conjecture in all dimensions, while the second part consists of a detailed analysis of the
eigenvalues coming from the different zones (resonance and non-resonance ones); when working in the resonance regions, we use some arguments from the theory of analytic functions of several complex variables. Dealing with the resonance regions is the
part of the proof which at the moment we cannot extend to higher dimensions (more on this later). Now let us discuss
the general strategy of the proof in detail (but still on a not too formal level)

The first step of the proof, as usual, consists of performing the Floquet-Bloch
decomposition to our operator \eqref{operator1}:
\bee\label{directintegral}
H=\int_{\oplus}H(\bk)d\bk,
\ene
where $H(\bk)=H_0+V(\bx)$
is the family of `twisted' operators with the same symbol as $H$ acting in
\bees
\GH:=L^2(\CO).
\enes
These auxiliary operators are labelled by
the quasi-momentum $\bk\in\CO^\dagger$;
the domain $\GD(\bk)$ of $H(\bk)$ consists of functions $f\in H^2(\CO)$
which are restrictions of functions $\hat f\in H^2_{loc}(\R^2)$ satisfying the following condition:
$\hat f(\bg +\bx)=e^{i\bk\bg}\hat f(\bx),\ \ \bg\in\Gamma$.
We refer the reader to \cite{ReeSim} for more details about this decomposition. Now it would be useful to introduce
a different density of states
\bee
N(\lambda):=\int_{\CO^{\dagger}}N(\la,H(\bk))d\bk
\ene
which is more convenient to deal with. It is known (see e.g. \cite{ReeSim}) that
\bee\label{tildeN}
\tilde N(\la)=\frac{1}{4\pi^2}N(\la).
\ene
Therefore, for our purposes it would be enough to prove \eqref{eq:intrnew1} for $N$ instead of $\tilde N$.

Note that we can assume without any loss of
generality that $\int_{\CO}V(\bx)d\bx=0$. Indeed, otherwise we consider a new operator $H_1:=H-b$ ($b$ is defined in \eqref{eq:b}).
Note that $H_1=-\Delta+V_1$ and the constant Fourier coefficient of $V_1:=V-b$ vanishes. Since
$N(\la;H)=N(\la-b,H_1)$, we see that asymptotic formulas \eqref{eq:intr2} for $H$ and $H_1$ are equivalent.
Therefore, we can (and will) always assume that $\int_{\CO}V(\bx)d\bx=0$.

Next, instead of trying to prove \eqref{eq:intrnew1} for all values of $\lambda$, we will prove it assuming that $\lambda$ is inside a fixed interval:
$\lambda\in [\lambda_n,16\lambda_n]$, where $\lambda_n=4^n\lambda_0$ is a large number, and we will allow the coefficients in \eqref{eq:intrnew1} to depend on $n$,
although the remainder should be uniform in $n$. In Section 3, we will show that if we can prove these asymptotic formulae for all $n$ with coefficients growing not too fast, this would imply the validity \eqref{eq:intrnew1} for all $\lambda$. The reason we require this reduction is the following: on later stages of the proof, we will decompose the
phase space (i.e. the space where the dual variable $\bxi$ lives)
into two regions: resonant and non-resonant zones. The resonant zones are, roughly speaking, the strips of some width $a$. The value of $a$ cannot be chosen the same for all values of $\lambda$, since we need $a$ to be of order $\la^{1/6}$. Thus, when we increase $\la$, at some stage we will have to increase the value
of $a$, and this can result in changes of the asymptotic coefficients in \eqref{eq:intrnew1}. However, if $\la$
runs over a fixed interval $[\lambda_n,16\lambda_n]$, we can keep $a$ fixed and thus the coefficients of our asymptotic
expansion \eqref{eq:intrnew1} will be fixed as well. Thus, starting from section 4, we will be assuming that
$\lambda\in [\lambda_n,16\lambda_n]$ and $n$ is fixed.

The next step is to assume that the potential $V$
is a finite trigonometric polynomial whose Fourier coefficients
\bee
\hat V(\bm):=\frac{1}{\sqrt{\vol(\CO)}}\int_{\CO}V(\bx)e^{-i\bm\bx}d\bx,\quad \bm\in\Gamma^{\dagger}
\ene
vanish when $|\bm|>R$. More precisely, we replace the original potential
\bee
V(\bx)=\frac{1}{\sqrt{\vol(\CO)}}\sum_{\bm\in
\Gamma^\dagger}\hat V(\bm)e^{-i\bm\bx}
\ene
by the truncated potential
\bee V'(\bx)=\frac{1}{\sqrt{\vol(\CO)}}\sum_{\bm\in
B(R)\cap\Gamma^\dagger}\hat V(\bm)e^{-i\bm\bx},
\ene
where $B(R)$ is a ball of radius $R$ centered at the origin.
Here, $R=R_n$ is a parameter which grows as a small positive power of $\lambda_n$ (for example,
$R_n=\la_n^{1/48}$).
It is easy to justify he fact that the error introduced by changing the potential in such a way is small;
this is where we use the fact that the original potential is infinitely smooth.
However, this truncation leaves us with an additional tedious job of checking how all the
important estimates depend on $R_n$.

Next, our aim is to construct a good approximation of all the eigenvalues of all operators $H(\bk)$ simultaneously (to be precise, we will need to approximate only eigenvalues which are inside the interval $[\lambda_n-100v,16\lambda_n+100v]$). 
In Section 4, we discuss what exactly
we mean by such a simultaneous approximation and prove that, if this approximation satisfies a bunch of additional
properties (in particular, the approximating function needs to behave in a proper way in a specially chosen coordinate system which should also satisfy certain properties), then asymptotic formula \eqref{eq:intrnew1} would follow automatically. This is done in Lemma
\ref{lem:4}. Unfortunately, we will not be able to use this lemma without modifications further on in the paper, but at least this lemma (and the proof of it) shows us which properties we are aiming for.

The main part of the paper, Sections 5--7, is devoted to the construction of an approximation of all the eigenvalues of all operators $H(\bk)$; the existence of such an approximation was assumed in Section 4. The main tool during
this construction will be an abstract result from perturbation theory -- Lemma \ref{perturbation2}.
This lemma allows us, under certain conditions, to study the spectrum of an operator $PHP$ instead of the spectrum
of an operator $H$. Here, $H=H_0+V$ is a bounded below operator with compact resolvent, $V$ is bounded, and $P$ is a spectral projection of $H_0$. Since the formulation of this lemma is rather involved, let us illustrate what it says by
considering a special case. Assume that $P$ is a further sum of spectral projections of $H_0$,
$P=\sum_{j=0}^J P_j$ such that the matrix of $V$ in the basis corresponding to $P_0,P_1,\dots,P_J,P_{J+1}:=I-P$ is block-three-diagonal (i.e. $P_jVP_t=0$ whenever $|j-t|>1$). Assume also that $\lambda=\lambda(H)$ is an eigenvalue of
$H$ and that the distance from the spectra of $P_jH_0P_j$ ($j=1,\dots,J+1$) to $\lambda$ is at least $a$, where $a$
is sufficiently large, so that $P_0HP_0$ is `essentially responsible' for the eigenvalue $\lambda$.
Then the operator $PHP$ has an eigenvalue $\lambda'$ such that $|\lambda-\lambda'|\ll a^{-2J}$. In applications, $a$ will
be of order $\lambda^{1/6}$, so by choosing
sufficiently large $J$, we can make our approximation as precise as we wish.

We are going to apply Lemma \ref{perturbation2} by constructing various projections $P$ such that the operator $PHP$
has an eigenvalue close to an eigenvalue of $H$. Roughly speaking, each point $\bxi$ from the phase space such that
$|\bxi|^2$ is close to $\la$ generates such a projection $P=P(\bxi)$. The structure of $P(\bxi)$ depends on the exact
location of $\bxi$ in the phase space. There are two types of points $\bxi$: resonant and non-resonant ones. For
non-resonant points $\bxi$, the structure of $P(\bxi)$ is relatively simple, the operator $PHP$ has a unique
eigenvalue close to $\lambda$, and we can find this eigenvalue using the standard approximating procedure
(for example, the Banach contraction mapping theorem).
Having constructed this approximation, we are ready to start computing $N(\la)$;
it is a relatively straightforward (but slightly tedious) task to compute the contribution to
the density of states coming from the non-resonant regions. The logarithmic terms appear on this stage (resonant regions do not produce
any logarithms).


In the case of the resonant $\bxi$, the structure of $P(\bxi)$ is more complicated, and therefore
it is much more difficult to compute a contribution to the density of states
coming from the resonance zones. The main problem lies in the fact that the approximation formula for the resonant eigenvalues
is not explicit: it expresses eigenvalues of $PHP$ in terms of the eigenvalues of an expression $A+\ep B$, where $A$ and $B$
are explicitly given symmetric matrices and $\ep\sim|\bxi|^{-1}$ is a small parameter which also depends on $\bxi$ in an explicit way. Of course, one can expand the eigenvalues of
$A+\ep B$ in powers of $\ep$, but the coefficients in this expansion will not be uniformly bounded in
$\bxi$,
so we will not be able to integrate this expansion in $\bxi$. Thus, we need to analyse the situation deeper. Let us denote by $\tilde P$
the projection onto the kernel of $A$. (We are interested in the perturbation of zero eigenvalues of $A$.) Then a priori there are two reasons why the coefficients in the asymptotic expansion of $A+\ep B$ can be large: either $A$ has eigenvalues close to zero (not the case in our situation), or the operator $\tilde PB\tilde P$ has eigenvalues close to each other. The latter possibility
is actually occurring in our problem. However, it turns out that $\tilde PB\tilde P$ is `essentially' unitary equivalent to a one-dimensional
Schr\"odinger operator with quasi-periodic boundary conditions on the interval. Therefore, there could be no more than two
eigenvalues of this operator located near each other (at this place we strongly use the fact that our operator $H$ is two-dimensional). The rest of the computations is similar to the non-resonance regions, only instead of solving equation
$\mu+G(\mu)=\lambda^{1/2}$ like we did in the non-resonance region (and where we used implicit function theorem), now we have
to solve the equation $\mu^2+X_1(\mu)\mu+X_2(\mu)=0$. The tool for dealing with equations of this type comes from the theory of functions of several complex variables and is called the Weierstrass Preparation Theorem. After using this theorem, we obtain the expressions for eigenvalues in the non-resonance regions; these expressions are no longer analytic in $\lambda^{1/2}$, but contain square
roots of analytic functions; however, these square roots will cancel after integration in $\bxi$ to produce an  asymptotic formula which contains only powers of
$\lambda^{1/2}$.

The rest of the paper is organised as follows: in the next section we give all necessary definitions and basic facts (the Weierstrass Preparation Theorem and corollaries from it). In Section 3 we reduce the problem of finding an asymptotic formula valid for all $\la$ to the problem of finding such a formula valid only for $\lambda$ inside a fixed interval. In Section 4
we describe what exactly we mean by a simultaneous approximation of all eigenvalues of all $H(\bk)$ and give some idea about the general strategy of the proof. In Section 5 we formulate auxiliary results which were proved in \cite{Par} and introduce the partition of the $\bxi$-plane into resonance and non-resonance regions. In Section 6 we deal with the non-resonance regions, and, finally, in Section 7 (the most complicated one) we compute the contribution to the density of states from the resonance zones.

{\bf Acknowledgment.} The first author acknowledges the warm hospitality of the Department of Mathematics
of University of Alabama at Birmingham where part of this work was carried out.
The work of the first author
was also partially supported by the Leverhulme trust and by the EPSRC grant EP/F029721/1.
The work of the second author was partially supported by the LMS grant. Both of us are very
grateful to Yu. Karpeshina and A. Sobolev for useful discussions. Finally, we thank the referees for
helpful comments.

\section{Notation and basic facts}

Let $\Gamma$ be a lattice in $\R^2$. We denote
by $\CO=\R^2/\Gamma$ the fundamental domain of $\Gamma$, by  
$\Gamma^{\dagger}$ the lattice dual to $\Gamma$, and by $\CO^{\dagger}=\R^2/\Gamma^{\dagger}$
its fundamental domain. 

For each vector $\bx\in\R^2$ we denote by $\bx^{\perp}$ the result of rotation of $\bx$ by $-\frac{\pi}{2}$
and $\bn(\bx):=\frac{\bx}{|\bx|}$, assuming $\bx\ne 0$. If $\bx_1,\bx_2\in\R^2$ are two non-zero vectors,
we denote by $\phi(\bx_1,\bx_2)$ the angle between them ($0\le\phi\le\pi$).

If $\bxi\in\R^2$, there exists unique decomposition $\bxi=\bg+\bk$ with $\bg\in\Gamma^\dagger$ and $\bk\in\CO^\dagger$. We call $\bg=:[\bxi]$ and $\bk=:\{\bxi\}$
resp. the {\it integer part}
and the {\it fractional part} of $\bxi$.

If $H$ is a bounded below self-adjoint operator with compact resolvent, then $\mu_j(H)$ is its $j$-th eigenvalue
(counting multiplicities).

We also assume that the average of $V$ over $\CO$ is zero.

By
$C$ or $c$ we denote positive constants,
The exact value of which can be different each time they
occur in the text,
possibly even each time they occur in the same formula. On the other hand, the constants which are labeled (like $C_1$, $c_3$, etc)
have their values being fixed throughout the text.
Given two positive functions $f$ and $g$,
we say that $f\gg g$, or $g\ll
f$, or $g=O(f)$ if the ratio $\frac{g}{f}$ is bounded. We say
$f\asymp g$ if $f\gg g$ and $f\ll g$.

The results in the rest of this section are quoted from \cite{Hor}.

\bet\label{WPT} (The Weierstrass preparation theorem). Let $F$ be
analytic and bounded in a neighborhood $\omega$ of $0$ in ${\mathbb
C}^n$ and assume that $F(0,z_n)/z_n^p$ is analytic and $\not=0$ at
$0$. (In other words it means that $(\partial^jF/\partial
z_n^j)(0,0)=0$ for $j=1,\dots,p-1,$ and $(\partial^pF/\partial
z_n^p)(0,0)\not=0$). Then one can find a polydisc ${\mathbf
D}\subset\omega$ such that every $G$ which is analytic and bounded
in ${\mathbf D}$ can be written in the form
\begin{equation}
G=qF+r,
\end{equation}
where $q$ and $r$ are analytic in ${\mathbf D}$, $r$ is a polynomial in $z_n$
of degree $<p$ (with coefficients depending on $z'=(z_1,\dots,z_{n-1})$)
and
\begin{equation}
\sup\limits_{\mathbf D} |q|\leq C\sup\limits_{\mathbf D} |G|.
\end{equation}
The representation is unique.
\ent

\ber It follows from the proof that polydisc ${\mathbf D}$
 and constant $C$ can be chosen to depend only on $\sup\limits_{\omega}
 |F|$, $p$ and $((\partial^pF/\partial z_n^p)(0,0))^{-1}$.
\enr

Now, choose $G:=z_n^p$ and put $W:=z_n^p-r,\ h:=q^{-1}$. We have
\bec\label{cor:WPT1} If $F$ satisfies the hypothesis of
Theorem~\ref{WPT}, then one can write $F$ in a unique way in the
form
\begin{equation}\label{representation}
F=hW,
\end{equation}
where $h$ and $W$ are
analytic in a neighborhood $\omega'$ of $0$, $h(0)\not=0$, and $W$ is a
Weierstrass polynomial, that is,
\begin{equation}
W(z)=z_n^p+\sum\limits_{j=0}^{p-1}a_j(z')z_n^j,
\end{equation}
where $a_j$ are analytic functions in a neighborhood of $0$ vanishing
when $z'=0$.

Moreover,
\begin{equation}\label{bound}
\sup\limits_{\omega'} (|h|+|h|^{-1})+\sup\limits_{\omega'} |W|\leq C_1,
\end{equation}
and $\omega'$ and $C_1$ depend only on $\sup\limits_{\omega}
 |F|$, $p$ and $((\partial^pF/\partial z_n^p)(0,0))^{-1}$.
\enc
\bec\label{cor:WPT2} Assume that a set of functions ${\mathbf
F}:=\{F\}$ satisfies the following properties:

1. Functions $F$ are analytic in a neighborhood $\omega$ of $0$.

2. For some $p$ we have $(\partial^jF/\partial z_n^j)(0,0)=0$ for $j=1,\dots,p-1,$ and
$(\partial^pF/\partial z_n^p)(0,0)\not=0$.

3. We have the bounds
\begin{equation}\label{unif}
\sup\limits_{F\in {\mathbf F}}\sup\limits_{\omega}
 |F|\leq C,\ \ \ \ \ \sup\limits_{F\in {\mathbf F}}|(\partial^pF/\partial
 z_n^p)(0,0)|^{-1}\leq C.
\end{equation}
Then there exist a neighborhood $\omega'$ and a constant $C_1$ such
that for any $F\in {\mathbf F}$ the representation
\eqref{representation} and estimate \eqref{bound} hold. Moreover,
$\omega'$ and $C_1$ are uniform with respect to $\{F\}$ and depend
only on $\omega,\ p$ and constant $C$ from \eqref{unif}.
\enc

\section{Reduction to a finite interval of spectral parameter}

The main result of our paper is the following theorem (or, rather, the corollary from it; we put
$\rho:=\sqrt{\la}$):

\bet\label{main_thm}
For each $K\in\N$ we have:
\bee\label{eq:main_thm1}
N(\rho^2)=\pi\rho^2+\sum_{j=0}^K e_j\rho^{-j}+\ln\rho\sum_{j=2}^K \hat e_j\rho^{-j}+o(\rho^{-K})
\ene
as $\rho\to\infty$. 
\ent
Once the theorem is proved, it immediately implies
\bec
For each $K\in\N$ we have:
\bee\label{eq:main_cor1}
\tilde N(\la)=\frac{1}{4\pi}\la-\frac{1}{4\pi|\CO|}\int_{\CO}V(\bx)d\bx+O(\la^{-K})
\ene
as $\la\to\infty$.
\enc
\bep
First of all, we notice that \cite{HitPol} implies that
\bee\label{Laplace}
\int_0^\infty e^{-t\la}\tilde N(\la)d\la\sim t^{-(d+2)/2}\sum_{l=0}^\infty q_jt^j
\ene
as $t\to 0+$, where $q_j$ are constants depending on the potential.
Now the corollary follows from theorem \ref{main_thm}, property \eqref{tildeN}, and calculations similar to that of \cite{KorPush}. Indeed, \cite{KorPush} implies that if all coefficients $\hat e_j$
vanish, then all coefficients $e_j,\ j>0$, vanish as well. It remains to show that all coefficients
$\hat e_j$ vanish. Suppose, this is not the case.
We consider separately even and odd values of $j$.  Suppose first that $\hat e_{2k}$ is the first non-zero even coefficient with hats. 
Then we consider the following integral:
\bee
I(t):=\int_1^\infty e^{-\la t}\la^{-k}\ln\la d\la
\ene
and, after elementary calculations, find that the asymptotic expansion of $I(t)$ as $t\to 0+$
contains a term $t^{k-1}\ln^2 t$ with a non-zero coefficient. This term is absent in the Laplace
transform of other terms from the expansion \eqref{eq:main_thm1}. Thus, our assumption that
$\hat e_{2k}\ne 0$ contradicts \eqref{Laplace}.

Suppose now that  $\hat e_{2k+1}$ is the first non-zero odd coefficient with hats. Then, similarly to the previous case, we consider the following integral:
\bee
I(t):=\int_1^\infty e^{-\la t}\la^{-(2k+1)/2}\ln\la d\la
\ene
and find that the asymptotic expansion of $I(t)$ as $t\to 0+$
contains a term $t^{(2k-1)/2}\ln t$ with a non-zero coefficient. This term is absent from the Laplace
transform of other terms from the expansion \eqref{eq:main_thm1}. Once again, we have reached a contradiction with \eqref{Laplace}.
Thus, all coefficients $\hat e_m$ vanish, and our corollary follows from \cite{KorPush}.


\enp

The rest of the paper is devoted to proving Theorem \ref{main_thm}.

To begin with, we choose sufficiently large $\rho_0>1$ (to be fixed later on) and put $\rho_n=2\rho_{n-1}=2^n\rho_0$;
we also define the interval
$I_n=[\rho_n,4\rho_n]$.
The proof of the main theorem will be based on the following lemma:

\bel\label{main_lem}
For each $M\in\N$ and $\rho\in I_n$ we have:
\bee\label{eq:main_lem1}
N(\rho)=\pi\rho^2+\sum_{j=0}^{6M}e_j(n)\rho^{-j}+\ln\rho\,\sum_{j=2}^{6M}\hat{e}_j(n)\rho^{-j}
+O(\rho_n^{-M}).
\ene
Here, $e_j(n),\ \hat{e}_j(n)$ are some real numbers depending on $j$ and $n$ (and $M$) satisfying
\bee\label{eq:main_lem2}
e_j(n)=O(\rho_n^{\frac{4j+7}{5}}),\ \ \hat{e}_j(n)=O(\rho_n^{\frac{2j+1}{3}}).
\ene
The constants in the $O$-terms do not depend on $n$ (but they may depend on $M$).
\enl
\ber\label{rem:new1}
Note that \eqref{eq:main_lem1} is not a `proper' asymptotic formula, since the coefficients
$e_j(n)$ are allowed to grow with $n$ (and, therefore, with $\rho$).
\enr

Let us prove theorem \ref{main_thm} assuming that we have proved lemma \ref{main_lem}. Let $M$ be fixed.
Denote
\bee
N_n(\rho^2):=\pi\rho^2+\sum_{j=0}^{6M}e_j(n)\rho^{-j}+\ln\rho\,\sum_{j=2}^{6M}\hat{e}_j(n)\rho^{-j}.
\ene
Then whenever $\rho\in J_n:=I_{n-1}\cap I_n=[\rho_n,2\rho_n]
$, we have:
\bee
N_n(\rho^2)-N_{n-1}(\rho^2)=\sum_{j=0}^{6M}t_j(n)\rho^{-j}+\ln\rho\,\sum_{j=2}^{6M}\hat{t}_j(n)\rho^{-j},
\ene
where
\bee
t_j(n):=e_j(n)-e_j(n-1),\qquad \hat t_j(n):=\hat e_j(n)-\hat e_j(n-1).
\ene
On the other hand, since for $\rho\in J_n$ we have both
$N(\rho)=N_n(\rho)+O(\rho_n^{-M})$ and $N(\rho)=N_{n-1}(\rho)+O(\rho_n^{-M})$, this implies that
$\sum_{j=0}^{6M}t_j(n)\rho^{-j}+\ln\rho\,\sum_{j=2}^{6M}\hat{t}_j(n)\rho^{-j}=O(\rho_n^{-M})$.
\becl
For each $j=0,\dots,6M$ we have:
$t_j(n)=O(\rho_n^{j-M}\ln\rho_n)$ and $\hat t_j(n)=O(\rho_n^{j-M})$.
\encl
\bep
Put $x:=\rho^{-1}$. Then $\sum_{j=0}^{6M}t_j(n)x^j-\ln x\,\sum_{j=2}^{6M}\hat{t}_j(n)x^{j}=O(\rho_n^{-M})$ whenever
$x\in [\frac{\rho_n^{-1}}{2},\rho_n^{-1}]$. Put $y:=x\rho_n$,
$\tau_j(n):=(t_j(n)+\hat t_j(n)\ln\rho_n)\rho_n^{M-j}$, and
$\hat \tau_j(n):=-\hat t_j(n)\rho_n^{M-j}$. Then
\bee\label{Cramer}
P(y):=\sum_{j=0}^{6M}\tau_j(n)y^j+\sum_{j=2}^{6M}\hat \tau_j(n)y^j\ln y=O(1)
\ene
whenever $y\in [\frac{1}{2},1]$. Consider the following $12 M$ functions: $y^j$ ($j=0,...,6M$)
and $y^j\ln y$ ($j=2,...,6M$) and label them $h_1(y),...h_{12M}(y)$. These functions are linearly independent on the interval
$[\frac{1}{2},1]$. Therefore, there exist points $y_1,...,y_{12M}\in [\frac{1}{2},1]$ such that
the determinant of the matrix $(h_j(y_l))_{j,l=1}^{12M}$ is non-zero. Now \eqref{Cramer} and the Cramer's Rule imply that for each $j$ the values $\tau_j(n)$ and $\hat\tau_j(n)$ are fractions with a bounded expression in the numerator and a fixed non-zero number in the denominator. Therefore,
$\tau_j(n)=O(1)$ and $\hat\tau_j(n)=O(1)$. This shows first that $\hat t_j(n)=O(\rho_n^{j-M})$ and then that $t_j(n)=O(\rho_n^{j-M}\ln\rho_n)$ as claimed.
\enp

Thus, for $j<M$, the series $\sum_{m=0}^\infty t_j(m)$ is absolutely convergent; moreover, for such
$j$ we have:
\bee
e_j(n)=e_j(0)+\sum_{m=1}^n t_j(m)=e_j(0)+\sum_{m=1}^\infty t_j(m)+O(\rho_n^{j-M}\ln\rho_n)=:e_j+O(\rho_n^{j-M}\ln\rho_n),
\ene
where we have denoted $e_j:=e_j(0)+\sum_{m=1}^\infty t_j(m)$. Similarly, for $j<M$ we have
\bee
\hat e_j(n)=\hat e_j(0)+\sum_{m=1}^n \hat t_j(m)=\hat e_j(0)+\sum_{m=1}^\infty \hat t_j(m)+O(\rho_n^{j-M})=:\hat e_j+O(\rho_n^{j-M}),
\ene
where we have denoted $\hat e_j:=\hat e_j(0)+\sum_{m=1}^\infty \hat t_j(m)$.

Since $e_j(n)=O(\rho_n^{\frac{4j+7}{5}})$
(it was one of the assumptions of lemma), we have:
\bee
\sum_{j=M}^{6M}|e_j(n)|\rho_n^{-j}=O(\rho_n^{\frac{7}{5}-\frac{M}{5}})=O(\rho_n^{-\frac{M}{6}}),
\ene
assuming as we can without loss of generality that $M$ is sufficiently large.
The sum with hats on is estimated similarly.
Thus, when $\rho\in I_n$, we have:
\bee\label{eq:main_lem11}
N(\rho)=\pi\rho^2+\sum_{j=0}^{M-1}e_j\rho^{-j}+\sum_{j=2}^{M-1}\hat e_j\rho^{-j}\ln\rho+O(\rho^{-M}\ln\rho)+O(\rho^{-\frac{M}{6}}).
\ene
Since constants in $O$ do not depend on $n$, 
for all $\rho\ge \rho_0$ we have:
\bee\label{eq:main_lem17}
\bes
N(\rho)&=\pi\rho^2+\sum_{j=0}^{M-1}e_j\rho^{-j}+\sum_{j=2}^{M-1}\hat e_j\rho^{-j}\ln\rho+O(\rho^{-\frac{M}{6}})\\
&=\pi\rho^2+\sum_{j=0}^{[M/6]}e_j\rho^{-j}+\sum_{j=2}^{[M/6]}\hat e_j\rho^{-j}\ln\rho+O(\rho^{-\frac{M}{6}}).
\end{split}
\ene
Taking $M=6K+1$, we obtain \eqref{eq:main_thm1}.

The rest of the paper is devoted to proving lemma \ref{main_lem}.

\section{Description of the approach. Integration in new coordinates}


From the previous section it is clear that we can study the density of states $N(\rho)$ assuming that
$\rho\in I_n$. Throughout the paper we will assume that $n$ is fixed and sometimes will omit index $n$ from the notation; however, we will carefully follow how all estimates depend on $n$. If we need to make sure that
$\rho_n$ is sufficiently large, we will achieve this by increasing $\rho_0$, keeping $n$ fixed.

First, we discuss the general strategy. In this section we describe how to construct the asymptotic formula
for $N(\rho)$ using certain objects (mappings $f$ and $g$ and coordinates $(r,\Phi)$ satisfying certain
properties); in the next sections, we will construct these objects.

Let us fix sufficiently large $n$, $\la=\rho^2$ with $\rho\in I_n$, and denote
\bee
\CA=\CA(\rho):=\{\bxi\in\R^2,\,|\bxi|^2\in [\la-100v,
\la+100v]\},
\ene
where $v:=||V||_{\infty}$. Obviously, $\CA$ is an annulus of width $\sim\rho^{-1}$. We also fix a number $M\in\N$.
Our aim is to construct good approximation of the eigenvalues lying close to $\la$. Namely, we will construct two mappings $f,g:\R^2\to\R$
such that for each $\bxi$, $f(\bxi)$ is an eigenvalue of $H(\{\bxi\})$; moreover, $f:\{\bxi\in\R^2,\,\{\bxi\}=\bk\}\to \sigma(H(\bk))$ is a bijection for each $\bk$ (here, we count all eigenvalues of $H(\bk)$ according to their multiplicities; the functions $f,g$ depend on $n$, $M$ and $\rho$). The difference $|f(\bxi)-g(\bxi)|$ is required to be sufficiently
small at least when $\bxi\in\CA$, namely, we postulate that the following two properties hold:

(i) $|f(\bxi)-g(\bxi)|\le\rho_{n}^{-M}$ for  $\bxi\in\CA$;

(ii) $|f(\bxi)-|\bxi^2||\le 2v$, similarly, $|g(\bxi)-|\bxi^2||\le 2v$.

Notice that the second property implies that
if $\bxi\not\in\CA$, then the following three inequalities are equivalent: $f(\bxi)<\la$ if and only if $g(\bxi)<\la$, and this in turn happens if and only if $|\bxi|<\rho$.

\ber
Rigorously speaking, the functions we will construct will satisfy property (i) not in the whole
annulus $\CA$, but in a slightly smaller annulus
$\{\bxi\in\R^2,\,|\bxi|^2\in [\la-90v,
\la+90v]\}$. Indeed, in the process of constructing $f$ and $g$ we will have to reduce
the width of the set $\CA$ by $2v$ several times. One obvious solution to this problem would be to introduce
sets $\CA_1=\{\bxi\in\R^2,\,|\bxi|^2\in [\la-98v,
\la+98v]\}$, $\CA_2$, etc. However, this would introduce extra notational complexity to a paper
which is already overburdened with notation. Thus, we will keep calling $\CA$ all annuli of slightly
smaller width whenever necessary.
\enr

Finally, we will construct function $g$ in such a way that it satisfies some asymptotic formulas. The next lemmas describe why these functions are going to be useful. Denote
\bees
B_f(\la):=f^{-1}((-\infty,\la]).
\enes

\bel\label{lem:1}
Suppose $f:\R^2\to\R$ is a measurable mapping such that $f:\{\bxi\in\R^2,\,\{\bxi\}=\bk\}\to \sigma(H(\bk))$ is a bijection (including multiplicities) for each $\bk$. Then
$N(\la)=\vol(B_f(\la))$.
\enl
\bep
Denote by $\chi_{B_f(\la)}$ the characteristic function of $B_f(\la)$. By Fubini's theorem we have:
\bee
\bes
\vol(B_f(\la))&=\int_{\R^2}\chi_{B_f(\la)}(\bxi)d\bxi\\&=\int_{\CO^{\dagger}}\#\{\bxi,\{\bxi\}=\bk\,\&\,f(\bxi)\le\la\}d\bk=
\int_{\CO^{\dagger}}N(\la,\bk)d\bk=N(\la).
\end{split}
\ene
\enp

Our next task is two-fold: to show that under certain conditions we can replace $B_f$ in lemma \ref{lem:1}
by $B_g$ so that the error is not too big and, secondly, to compute $\vol(B_g(\la))$ (or, at least, to
expand this volume in powers of $\la$). Unfortunately, assumptions (i) and (ii) on functions $f$ and $g$ made above are not the only necessary requirements to do this job: we also need to check that function $g$ behaves in a `nice' way in some suitable coordinates. Since the complete set of required conditions
looks rather nasty, we will introduce these conditions slowly, on at a time, to show why each particular condition is required. First, we check that the polar coordinates could do the trick.

\bel\label{lem:2}
Let $\lambda=\rho^2$ be fixed and let $N$ be a fixed natural number.
Suppose $f:\R^2\to\R$ is a measurable mapping such that $f:\{\bxi\in\R^2,\,\{\bxi\}=\bk\}\to \sigma(H(\bk))$ is a bijection for each $\bk$ (counting multiplicities).
Suppose, $g:\R^2\to\R$
is a measurable mapping and that $f,g$ satisfy properties (i) and (ii) above.
Suppose also that
$\frac{\partial g}{\partial r}(re^{i\phi})\gg\rho$ whenever $\bxi=re^{i\phi}\in\CA$.
Then $N(\la)=\vol(B_g(\la))+O(\rho^{-M})$.
\enl
\bep
Assumptions of lemma (namely, property (ii) above) imply that the symmetric difference $B_f(\la)\triangle B_g(\la)\subset\CA$.
The boundary of the `ball' $B_g(\la)$ is a subset of $\CA$; since the function $g=g(r,\phi)$ is increasing in $r$, for any fixed $\phi_0$ the intersection of $B_g(\la)$ with any semi-infinite interval
$\{re^{i\phi_0},\,r\in [0,\infty)\}$
is an interval $\{re^{i\phi_0},\,r\in [0,Z]\}$, where $Z=Z(\phi_0)$ is a well-defined function. Since
$\frac{\partial g}{\partial r}\ge C_1\rho$, we also have that if $r<Z(\phi)-C_1^{-1}\rho^{-M-1}$, then
$\la-g(re^{i\phi})>\rho^{-M}$, and so $f(re^{i\phi})<\lambda$ and
$\bxi=r e^{i\phi}\in B_f(\la)$. Similarly, if
$r>Z(\phi)+C_1^{-1}\rho^{-M-1}$, then
$g(re^{i\phi})-\la>\rho^{-M}$, and so $f(re^{i\phi})>\lambda$ and $\bxi=r e^{i\phi}\not\in B_f(\la)$. Thus, the symmetric difference
$B_f(\la)\triangle B_g(\la)\subset\{\bxi=r e^{i\phi}\in\R^2,\,|r-Z(\phi)|\le C_1^{-1}\rho^{-M-1}\}$ and thus
$\vol(B_f(\la)\triangle B_g(\la))\ll\rho^{-M}$. Together with lemma \ref{lem:1}, this finishes the proof.
\enp
Later on, we will apply lemma \ref{lem:2} in a more general situation, when $r$ is not precisely
the radial coordinate, but `close' to the radial coordinate in a certain sense; more precisely, we will need the
following statement
(with proof 
being exactly the same as proof of lemma \ref{lem:2}):
\bec\label{cor:3}
Let $\CS$ be a curve of length $\ll 1$, and let $(r,\Phi)$ ($r\in\R^+$, $\Phi\in\CS$) be coordinates in $\CA(\rho)$ such that the Jacobian
$\bigm|\frac{\partial(x,y)}{\partial(r,\Phi)}\bigm|\ll \rho$. Suppose, $f:\R^2\to\R$ is a measurable mapping such that $f:\{\bxi\in\R^2,\,\{\bxi\}=\bk\}\to \sigma(H(\bk))$ is a bijection for each $\bk$ (counting multiplicities). 
Suppose, $g:\R^2\to\R$
is a measurable mapping and that $f,g$ satisfy properties (i) and (ii) above.
Suppose also that
$\frac{\partial g}{\partial r}(r,\Phi)\gg\rho$.
Then $N(\la)=\vol(B_g(\la))+O(\rho^{-M})$.
\enc
\ber
Obviously, lemma \ref{lem:2} is a special case of corollary \ref{cor:3} with $\CS= S^{1}$ being a circle of radius $1$ centered at the origin and $(r,\Phi)$ being the usual polar coordinates.
\enr
\ber\label{rem:alternative}
Suppose that another set of coordinates $(\tilde r,\tilde \Phi)$ satisfy slightly different conditions:
$\tilde \Phi\in\tilde\CS$, where $\tilde\CS$ is a curse of length $\ll \rho$, but the Jacobian
$\bigm|\frac{\partial(x,y)}{\partial(\tilde r,\tilde\Phi)}\bigm|\ll 1$. Then the coordinates
$(r,\Phi):=(\tilde r,\frac{\tilde\Phi}{\rho})$ satisfy all assumptions of corollary \ref{cor:3}, so the
conclusion of this corollary will also be valid for such coordinates. We will be using both types of coordinates, depending upon convenience.
\enr
\ber\label{assumptions1}
The coordinates $(r,\Phi)$ which we will introduce in further sections will be defined simultaneously for all $\rho\in I_n$, i.e. they will be defined for all points
\bee\label{CA_n}
\bxi\in\CA^{(n)}:=\cup_{\rho\in I_n}\CA(\rho).
\ene
Unfortunately, we need to make our assumptions about the coordinate system $(r,\Phi)$ even more complicated.
First of all, we will need to use different coordinates systems in different parts of $\CA^{(n)}$, so we assume
that we have a decomposition of $\CA^{(n)}$ as a disjoint union:
\bee\label{sqcup}
\CA^{(n)}=\sqcup_{l=1}^L \CA^{(n)}_l;
\ene
for simplicity,
we assume that all sets $\CA^{(n)}_l$ are open, and treat \eqref{sqcup} modulo points
on the boundaries of these sets. We also assume that there is a coordinate system $(r,\Phi)$ in each
$\CA^{(n)}_l$ ($r(\bxi)\in\R^+$, $\Phi(\bxi)\in\CS_l$, where $\CS_l$ is a curve of length $\ll 1$) and that this system
satisfies all assumptions of corollary \ref{cor:3}. Whenever we talk about the Jacobian $\bigm|\frac{\partial(x,y)}{\partial(r,\Phi)}\bigm|$, we will assume that it is defined only at points $\bxi$
located inside some $\CA^{(n)}_l$, i.e. the Jacobian is not defined for points on the boundary of $\CA^{(n)}_l$.
Other conditions we always assume are: $r(\bxi)\sim|\bxi|$, and for each fixed $\Phi_0$ the
intersection
\bee\label{eq:r1}
\CA^{(n)}_l\cap\{\bxi=(r,\Phi_0),\,r\in [0,\infty)\}=\{\bxi=(r,\Phi_0),\,r\in [r_1,r_2]\}
\ene
is an interval with endpoints $\bxi_1=(r_1,\Phi_0)$ and $\bxi_2=(r_2,\Phi_0)$ satisfying $|\bxi_1|^2=\rho_n^2-100v$ and
$|\bxi_2|^2=(4\rho_n)^2+100v$. The latter condition, while looking rather horrific, is easy to check
and will be always automatically satisfied in our constructions. Roughly speaking, it is needed to ensure that
the curve $\{\bxi=(r,\Phi_0),\,r\in [0,\infty)\}$ (which happens to be a semi-infinite interval in all our constructions) cannot enter or leave $\CA^{(n)}_l$ from the `sides'. Technically, it is required to make sure
that formulas \eqref{eq:414} and \eqref{eq:415} imply \eqref{eq:416}.
\enr
Let us introduce more notation. Put
\bee\label{eq:47}
\hat A^+:=\{\bxi\in\R^2,\,g(\bxi)<\rho^2<|\bxi|^2\}
\ene
and
\bee\label{eq:48}
\hat A^-:=\{\bxi\in\R^2,\,|\bxi|^2<\rho^2<g(\bxi)\}.
\ene
\bel\label{lem:new1}
\bee\label{eq:46}
\vol(B_g(\rho^2))=\pi\rho^2+\vol\hat A^+-\vol\hat A^-.
\ene
\enl
\bep
We obviously have $B_g(\rho^2)=B(\rho^2)\cup \hat A^+\setminus \hat A^-$. Since $\hat A^-\subset B(\rho^2)$
and $\hat A^+\cap B(\rho^2)=\emptyset$, this implies \eqref{eq:46}.
\enp
\ber\label{rem:nnew1}
Property (ii) of the mapping $g$ implies that we have $\hat A^+,\ \hat A^-\subset\CA$. Thus, statements \ref{lem:1}--\ref{lem:new1}
imply that in order to compute $N(\lambda)$, we need to analyse the behaviour of $g$ only inside $\CA$.
\enr
In order to apply corollary \ref{cor:3} and lemma \ref{lem:new1} for computing the asymptotic behaviour of $N(\rho)$, we need even more assumptions. Roughly speaking, the next lemma says that whenever all objects involved enjoy a power asymptotics
at infinity, then so does $B_g(\la)$.
\bel\label{lem:4}
Let $l\in{\mathbb N}$ and $\alpha\in(0,1)$ be fixed.
Suppose 
that all assumptions of corollary \ref{cor:3} and remark \ref{assumptions1} are satisfied and
that for fixed $\Phi$ the point $\bxi=(r,\Phi)\in\CA^{(n)}_l$ has an absolute value $|\bxi|$ which has an asymptotic expansion in powers of $r$:
\bee\label{eq:41}
|\bxi|=r(\bxi)\left(1+\sum_{j=1}^{\left[\frac{M+1}{1-\alpha}\right]} a_j(\Phi(\bxi))r(\bxi)^{-j}\right)+O(r(\bxi)^{-M}).
\ene
and this formula can be formally differentiated once with respect to $r$, i.e.
\bee\label{eq:41n}
\frac{\partial |\bxi|}{\partial r}=1-\sum_{j=2}^{\left[\frac{M+1}{1-\alpha}\right]} (j-1) a_j(\Phi(\bxi))r(\bxi)^{-j}+O(r(\bxi)^{-M-1}).
\ene
Suppose also that the function $g$ enjoys the following asymptotic behaviour in $r(\bxi)$ when $\bxi\in\CA^{(n)}_l$:
\bee\label{eq:45}
g(\bxi)= r(\bxi)^2\left(1+\sum_{j=1}^{\left[\frac{M+2}{1-\alpha}\right]} \check a_j(\Phi(\bxi))r(\bxi)^{-j}\right)+O(r(\bxi)^{-M}).
\ene
Finally, suppose that the Jacobian also satisfies an asymptotic formula:
\bee\label{eq:43}
\frac{\partial(x,y)}{\partial(r,\Phi)}= r(\bxi)+\sum_{j=1}^{\left[\frac{M}{1-\alpha}\right]} \hat a_j(\Phi(\bxi))r(\bxi)^{-j}+O(r(\bxi)^{-M}).
\ene
All functions $a_j$, $\check a_j$, etc. are measurable and bounded (but not necessarily continuous) functions of $\Phi$ and
are $O(\rho^{\al j})$, $\al<1$. Then
\bee\label{eq:44}
\vol(\hat A^+\cap\CA^{(n)}_l)-\vol(\hat A^-\cap\CA^{(n)}_l)=\rho^2\sum_{j=1}^{\left[\frac{M+2}{1-\alpha}\right]}
b_j \rho^{-j}+O(\rho^{-M})
\ene
and all $b_j$ are $O(\rho^{\al j})$.
\enl
\ber
It may seem strange that absolute value of the power in the remainder term in the above formulas is smaller than the upper summation limit. This is caused by the fact that the coefficients $a_j$, $\check a_j$, $b_j$, etc. are allowed to grow together with $\rho$: compare this with remark \ref{rem:new1}.
\enr
\bep

First of all we notice that without loss of generality we can assume that
\bee\label{eq:45n}
g(\bxi)= r(\bxi)^2\left(1+\sum_{j=1}^{\left[\frac{M+2}{1-\alpha}\right]} \check a_j(\Phi(\bxi))r(\bxi)^{-j}\right),
\ene
since corollary \ref{cor:3} implies that the error caused by using this approximation is $O(\rho^{-M})$.
Let us for a moment fix some value $\Phi_0$. 
Then 
the RHS of \eqref{eq:45n}
is an increasing function of $r$ for sufficiently large $r$. Let us call by $Q^1=Q^1_{\Phi}$ the inverse function to \eqref{eq:45n}, i.e.
\bee\label{eq:410}
(Q^1(t))^2\left(1+\sum_{j=1}^{\left[\frac{M+2}{1-\alpha}\right]} \check a_j(\Phi(\bxi))(Q^1(t))^{-j}\right)
=t.
\ene
It is an easy exercise to show that the function $Q^1$ also enjoys the asymptotic behaviour as $|\bxi|\to\infty$:
\bee\label{eq:411}
Q^1_{\Phi}(t)= t^{1/2}\left(1+\sum_{j=1}^{\left[\frac{M+1}{1-\alpha}\right]} \check b_j(\Phi(\bxi))t^{-j/2}\right)+O(t^{-M/2})
\ene
and that the coefficients $\check b_j=O(\rho^{\al j})$.
Note that $Q^1$ is also monotone increasing, so the inequality
$g(\bxi)<\rho^2$ is equivalent to $r(\bxi)<Q^1_{\Phi}(\rho^2)$.

Equation \eqref{eq:41n} implies that the RHS of \eqref{eq:41} is an increasing function of $r$.
Let us denote by $Q^2=Q^2_{\Phi}$ the inverse function to it. Then again it is easy to show
that $Q^2$ also enjoys the asymptotic behaviour:
\bee\label{eq:412}
Q^2(t)=t\left(1+\sum_{j=1}^{\left[\frac{M+1}{1-\alpha}\right]} \hat b_j(\Phi(\bxi))t^{-j}\right)+O(t^{-M})
\ene
with $\hat b_j=O(\rho^{\alpha j})$.
Moreover, $Q^2(t)$ is a monotone function for large $t$, so the inequality
$|\bxi|<\rho$ is equivalent to $r(\bxi)<Q^2_{\Phi}(\rho)$.

Now we can re-write definitions \eqref{eq:47}-\eqref{eq:48} in the following way:
\bee\label{eq:414}
\hat A^+:=\{\bxi\in\R^2,\,Q^2_{\Phi}(\rho)<r(\bxi)<Q^1_{\Phi}(\rho^2)\}
\ene
and
\bee\label{eq:415}
\hat A^-:=\{\bxi\in\R^2,\,Q^1_{\Phi}(\rho^2)<r(\bxi)<Q^2_{\Phi}(\rho)\}.
\ene
Therefore,
\bee\label{eq:416}
\vol(\hat A^+\cap\CA^{(n)}_l)-\vol(\hat A^-\cap\CA^{(n)}_l)
=\int_{\CS_l}\int^{Q^1_\Phi(\rho^2)}_{Q^2_\Phi(\rho)}\frac{\partial(x,y)}{\partial(r,\Phi)}dr d\Phi.
\ene
Now \eqref{eq:44} follows from \eqref{eq:411}, \eqref{eq:412}, and \eqref{eq:43}.
\enp
\ber
When applying lemma \ref{lem:4} later, we will first establish asymptotic formula \eqref{eq:45} only for $\bxi\in\CA(\rho)$
with a fixed $\rho\in I_n$. After this formula is established for each $\rho\in I_n$, we just check that the
coefficients do not depend on the particular choice of $\rho$, so this formula holds for all $\bxi\in\CA^{(n)}_l$.
\enr
\ber
Note that logarithms have made a brief appearance in the RHS of \eqref{eq:416} before being
canceled out.
\enr
\ber
Lemma~\ref{lem:4} gives us only a priori estimates on
coefficients $b_j$. In fact, we will be able to say more about them. For example, since $\hat A^+,\ \hat A^-\subset\CA$
and $\vol\CA\ll 1$, this implies that the LHS of \eqref{eq:44} is
bounded and, thus, leads to additional restrictions on the first several coefficients $b_j$. Later we will come back to this discussion.
\enr

\section{Abstract perturbation results and decomposition into invariant subspaces}

In this section, we begin the construction of the mappings $f$, $g$ with properties (i), (ii) stated in the
previous section.

First, we formulate the abstract result which was proved in \cite{Par} (lemma 3.2 and corollary 3.3);
see introduction for an informal discussion of this result.

\bel\label{perturbation2} Let $H_0$, and $V$ be self-adjoint
operators such that $H_0$ is bounded below and has compact
resolvent and $V$ is bounded. Let $\{P^l\}$ ($l=0,\dots,L$) be a
collection of orthogonal projections commuting with $H_0$ such
that if $l\ne n$ then $P^lP^n=P^lVP^n=0$. Denote $Q:=I-\sum P^l$.
Suppose that each $P^l$ is a further sum of orthogonal projections
commuting with $H_0$: $P^l=\sum_{j=0}^{j_l} P^l_j$ such that
$P_j^lVP_t^l=0$ for $|j-t|>1$ and $P_j^lVQ=0$ if $j<j_l$. Let
$v:=||V||$ and let us fix an interval $\BJ=[\la_1,\la_2]$ on the
spectral axis which satisfies the following properties: spectra of
the operators $QH_0Q$ and $P_j^lH_0P_j^l$, $j\ge 1$ lie outside $\BJ$;
moreover, the distance from the spectrum of $QH_0Q$ to $\BJ$ is
greater than $4v$ and the distance from the spectrum of
$P_j^l H_0P_j^l$ ($j\ge 1$) to $\BJ$, which we denote by $a_j^l$, is
greater than $12v$. Denote by $\mu_p\le\dots\le\mu_q$ all
eigenvalues of $H=H_0+V$ which are inside $\BJ$. Then the
corresponding eigenvalues $\tilde\mu_p,\dots,\tilde\mu_q$ of the
operator
\bees
\tilde H:=\sum_l P^lHP^l+QH_0Q
\enes
are eigenvalues of
$\sum_l P^lHP^l$, and they satisfy
\bees
|\tilde\mu_r-\mu_r|\le
\max_l\left[(6v)^{2j_l+1}\prod_{j=1}^{j_l}(a_j^l-6v)^{-2}\right];
\enes
all other
eigenvalues of $\tilde H$ are outside the interval
$[\la_1+v,\la_2-v]$. More precisely, there exists an injection
$G$ defined on the set of eigenvalues of the operator $\sum_l P^lHP^l$ (all eigenvalues are
counted according to their multiplicities) and mapping them to the subset of the
set of eigenvalues of
$H$ (again considered counting multiplicities) such that:

(a) all eigenvalues of $H$ inside $\BJ$ have a pre-image,

(b) If $\hat\mu_r\in [\la_1+2v,\la_2-2v]$ is an eigenvalue of $\sum_l P^lHP^l$, then
\bees
|G(\hat\mu_r)-\mu_r|\le
\max_l \left[(6v)^{2j_l+1}\prod_{j=1}^{j_l}(a_j^l-6v)^{-2}\right],
\enes

and

(c) $G(\hat\mu_r)=\mu_{r+T}(H)$, where $T$ is the number of eigenvalues of
$QH_0Q$ which are smaller than $\lambda_1$.

Finally, we have: $||H-\tilde H||\le 2v$.
\enl

Let us fix $n$ and $M$, and let $\la=\rho^2$ be a real number with $\rho\in I_n$. Consider the truncated potential
\bee V'(\bx)=\sum_{\bm\in
B(R_n)\cap\Gamma^\dagger}\hat V(\bm)\be_{\bm}(\bx),
\ene
where
\begin{equation*}
\be_{\bm}(\bx):=\frac{1}{\sqrt{\vol(\CO)}} e^{i\lu\bm,\bx\ru},\ \ \bm \in
\Gamma^\dagger,
\end{equation*}
and
\bee\label{Fourier}
\hat V(\bm)=\int_{\CO}
V(\bx)\be_{-\bm}(\bx)d\bx
\ene
are the Fourier coefficients of $V$. $R_n$
is a large parameter the precise value of which will be chosen
later; at the moment we just state that $R_n\sim \rho_n^{p}$ with
$p>0$ being small. Throughout the text, we will prove various
statements which will hold under conditions of the type $R_n<\rho_n^{p_j}$.
After each statement of this type, we will always assume, without possibly specifically
mentioning, that these conditions are always satisfied in what follows; at the end, we will
choose $p=\min p_j$.

Since $V$ is smooth, for each $m$ we have
\bee
\sup_{\bx\in\R^2}|V(\bx)-V'(\bx)|\ll R_n^{-m}.
\ene
This implies
that if we denote $H'(\bk):=H_0(\bk)+V'$ with the domain
$\CD(\bk)$, the following
estimate holds for all $n$:
\bee\label{truncate1}
|\mu_j(H(\bk))-\mu_j(H'(\bk))|\ll R_n^{-m}\ll\rho_n^{-mp}.
\ene
Thus, if we choose sufficiently large $m$, namely $m>M/p$, we can safely work with the truncated operator
$H'$ instead of the original operator $H$.

For each natural $j$
we denote
\bee\label{Tj}
\T_j:=\Gamma^\dagger\cap B(jR_n),\, \T_0:=\{0\}, \, \T'_j:=\T_j\setminus \{0\}.
\ene

We also choose a number $\tilde M:=3 M$. Each vector from $\bg\in\T'_{6\tilde M}$ generates a one-dimensional linear space $\{t\bg,\,t\in\R\}$.
The intersection  $\{t\bg,\,t\in\R\}\cap\T'_{6\tilde M}$ contains two vectors with the smallest length.
We call such vectors the {\it primitive} vectors. Note that if $\bth$ is a primitive vector, then
so is $-\bth$.
Let $\bth_1,\dots,\bth_L$ be the set of all the primitive elements of $\T'_{6\tilde M}$. We choose the labeling in
such a way that if we take $\bn(\bth_1)$ and start rotating it counterclockwise, we meet $\bn(\bth_2)$, $\bn(\bth_3)$,
etc. in consecutive order.
\bel\label{angle}
If $\bg,\ \bnu\in\T'_{15\tilde M}$ are two linearly independent vectors, then
the angle $\phi(\bg,\bnu)\gg R_n^{-2}$ for large $R_n$.
\enl
\bep
It is a simple geometry (and was proved, e.g. in \cite{Par}, lemma 4.2 and corollary
4.3).
\enp
\bec\label{angle1}
Under assumptions of lemma \ref{angle} we have $|\lu\bn(\bg),\bn(\bnu^{\perp})\ru|\gg R_n^{-2}$ for large $R_n$.
\enc

Let $\bth=\bth_l$ be a primitive vector which we consider fixed for the moment. Let us introduce cartesian coordinates
on a plane where the first axis goes along $\bth^{\perp}$, and the second axis goes along $\bth$. We call this set of coordinates
{\it coordinates generated by} $\bth$. Sometimes, we will also need the cartesian coordinates which are fixed and independent of
the choice of $\bth_l$; we will call such set of coordinates {\it universal coordinates}.

This choice of coordinates generated by $\bth$ means
that each $\bxi\in\R^2$ has coordinates $(\xi_1,\xi_2)$, where $\xi_1=\lu\bxi,\bn(\bth^{\perp})\ru$ and
$\xi_2=\lu\bxi,\bn(\bth)\ru$. Let us fix this coordinate system for now.
We also define $a=a_n$ to be the smallest real number which satisfies two conditions:
\bee\label{conditionsa}
a\ge\rho_n^{1/3} \quad\&\quad \frac{2a}{|\bth|}-\frac{1}{2}\in\N.
\ene
In particular, we have $\rho_n^{1/3}\leq a\leq \rho_n^{1/3} +|\bth|/2 +1\leq 2\rho_n^{1/3}$. Now we can make the following definitions:
\bee\label{L}
\L(\bth):=\{\bxi\in\R^2,\, |\lu\bxi,\bn(\bth)\ru|<a\},
\ene
\bee\label{Xi1}
\Xi_1(\bth):=\{\bxi\in\CA(\rho)\cap\L(\bth),\,\, \lu \bxi,\bth^{\perp}\ru >0\}.
\ene
Obviously, the intersection $\CA(\rho)\cap\L(\bth)$ consists of two connected components, and the condition
$\lu \bxi,\bth^{\perp}\ru >0$ chooses one of them. We also define
\bee\label{Xi2}
\Xi_2(\bth):=\{\boldeta=\bxi+t\bth,\, \bxi\in\Xi_1(\bth),\,t\in\R\},
\ene
\bee\label{Xi3}
\Xi_3(\bth):=\Xi_2(\bth)\cap\L(\bth),
\ene
and
\bee\label{Xi4}
\Xi_4(\bth):=\bigl(\CA(\rho)\cap\Xi_2(\bth)\bigr)\setminus \Xi_3(\bth).
\ene

\begin{figure}[htb]
    \centering
    \psfrag{a}[][]{$a$}
    \psfrag{-a}[][]{$-a$}
    \psfrag{L}[][]{$\L(\bth)$}
    \psfrag{Ar}[][]{$\CA(\rho)$}
    \psfrag{t}[][]{$\bth$}
    \psfrag{p}[][]{$\bth^{\perp}$}
    \psfrag{X1}[][]{$\Xi_1(\bth)$}
    \includegraphics[width=3in]{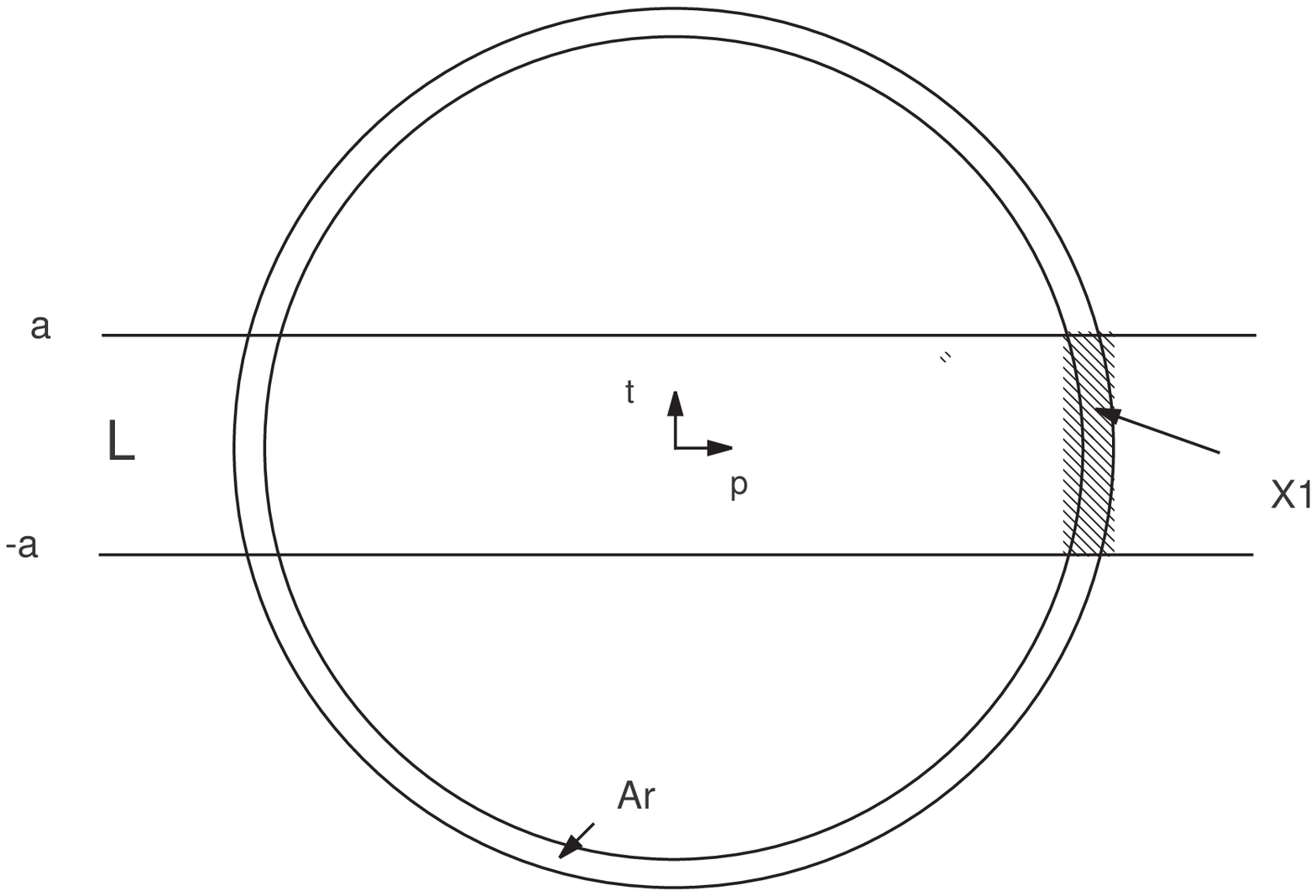}
    \centerline{Figure 1}\label{Fig1}
\end{figure}
\begin{figure}[htb]
    \centering
    \psfrag{a}[][]{$a$}
    \psfrag{-a}[][]{$-a$}
    \psfrag{L}[][]{$\L(\bth)$}
    \psfrag{Ar}[][]{$\CA(\rho)$}
    \psfrag{t}[][]{$\bth$}
    \psfrag{p}[][]{$\bth^{\perp}$}
    \psfrag{X2}[][]{$\Xi_2(\bth)$}
    \includegraphics[width=3in]{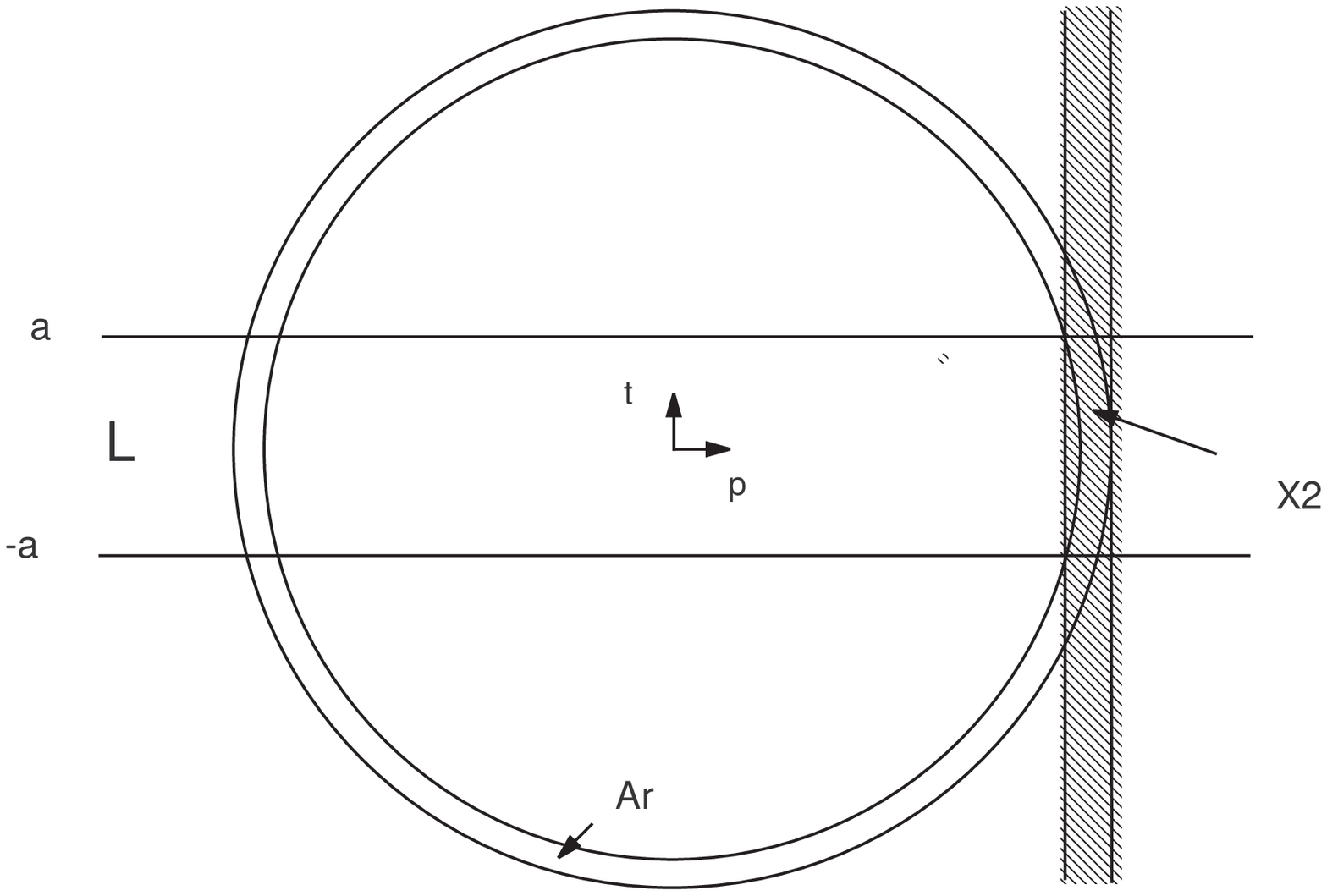}
    \centerline{Figure 2}\label{Fig2}
\end{figure}
\begin{figure}[htb]
    \centering
    \psfrag{a}[][]{$a$}
    \psfrag{-a}[][]{$-a$}
    \psfrag{L}[][]{$\L(\bth)$}
    \psfrag{Ar}[][]{$\CA(\rho)$}
    \psfrag{t}[][]{$\bth$}
    \psfrag{p}[][]{$\bth^{\perp}$}
    \psfrag{X3}[][]{$\Xi_3(\bth)$}
    \psfrag{X4}[][]{$\Xi_4(\bth)$}
    \includegraphics[width=3in]{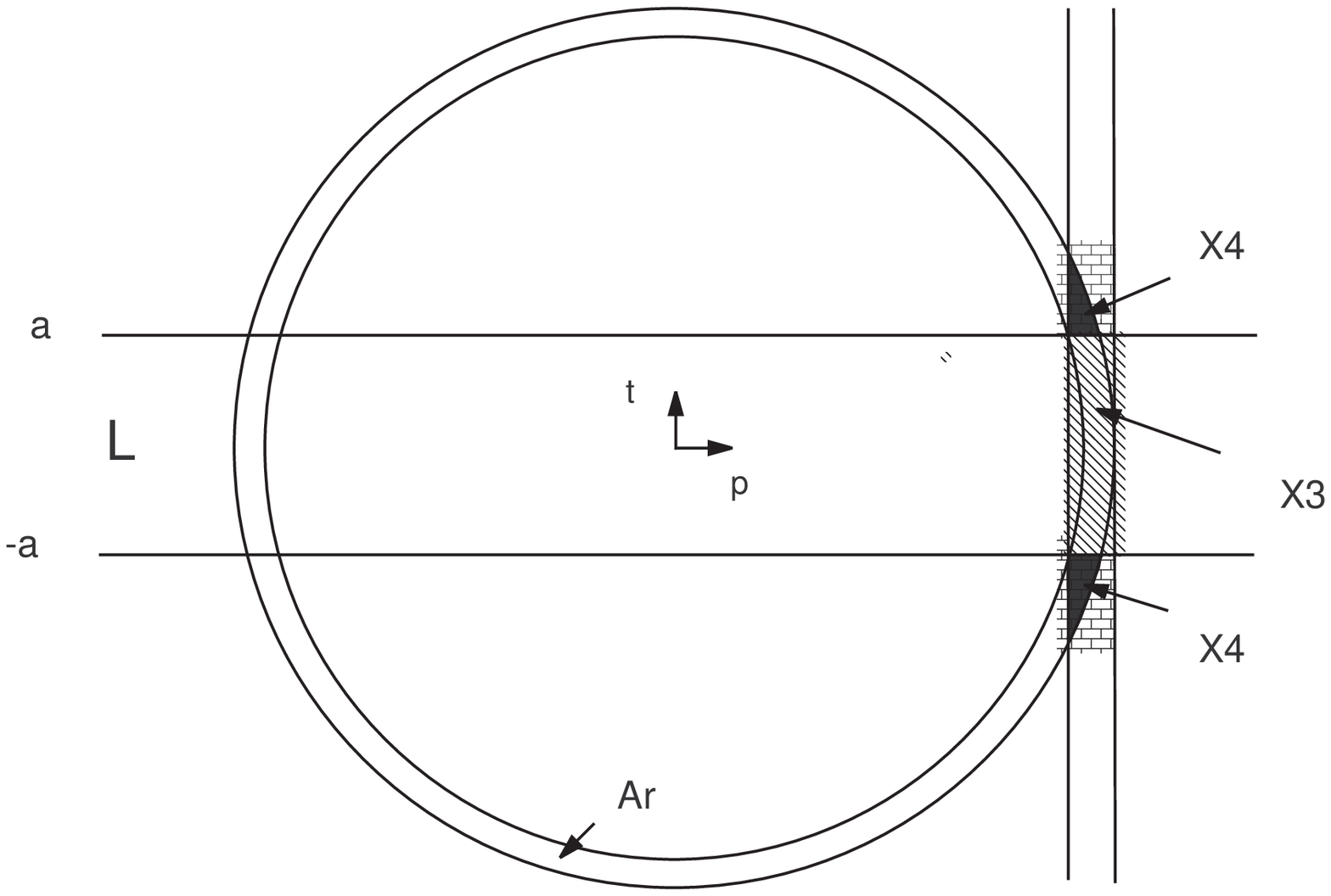}
    \centerline{Figure 3}\label{Fig3}
\end{figure}

\vskip 2cm
\bel\label{lem:Xi1}
Suppose $R_n\le\rho_n^{1/10}$, $\bxi\not\in \L(\bth)$ and $\bg=t\bth\in\T'_{6\tilde M}$. Then
$||\bxi+\bg|^2-|\bxi|^2|\gg\rho_n^{1/3}$.
\enl
\bep
Indeed, we have
\bees
||\bxi+\bg|^2-|\bxi|^2|\ge |\lu\bxi,\bg\ru|-|\bg|^2
\gg a\ge\rho_n^{1/3}.
\enes
\enp
\bel\label{lem:Xi2}
Let $\boldeta\in\Xi_1(\bth)$. Then $|\eta_1-\rho|\ll\rho^{-1/3}$.
\enl
\bep
Indeed, since $\boldeta\in\CA(\rho)$, we have $|\boldeta|^2=\eta_1^2+\eta_2^2=\rho^2+O(1)$. However, since
$\boldeta\in\L(\bth)$, we have $\eta_2^2=O(\rho^{2/3})$. Thus,
\bees
\eta_1=(\rho^2+O(\rho^{2/3}))^{1/2}=\rho(1+O(\rho^{-4/3}))^{1/2}=\rho(1+O(\rho^{-4/3}))=\rho+O(\rho^{-1/3})),
\enes
which finishes the proof (recall that $\eta_1$ is positive).
\enp
Since points in $\Xi_2$ have the same first coordinate as the points from $\Xi_1$, we immediately obtain:
\bec\label{cor:Xi3}
Let $\bxi\in\Xi_2(\bth)$. Then $|\xi_1-\rho|\ll\rho^{-1/3}$.
\enc
Let us denote
\bees
p_-=p_-(\bth):=\inf\{\eta_1,\,\boldeta=(\eta_1,\eta_2)\in\Xi_1(\bth)\}
\enes
and
\bees
p_+=p_+(\bth):=\sup\{\eta_1,\,\boldeta=(\eta_1,\eta_2)\in\Xi_1(\bth)\}.
\enes
Then lemma \ref{lem:Xi2} implies $p_+-p_-\ll \rho^{-1/3}$. Moreover, we can give another equivalent definition
of $\Xi_2$:
\bee\label{Xi21}
\Xi_2(\bth)=\{\boldeta=(\eta_1,\eta_2),\,\eta_1\in (p_-,p_+)\}.
\ene
Note that we obviously have the following equalities:
\bee\label{p-}
p_-=\inf\{\eta_1,\,\boldeta=(\eta_1,\eta_2)\in\Xi_4(\bth)\}
\ene
and
\bee\label{a}
a=\inf\{|\eta_2|,\,\boldeta=(\eta_1,\eta_2)\in\Xi_4(\bth)\}.
\ene
Denote
\bee\label{tildep-}
\tilde p_-:=\sup\{\eta_1,\,\boldeta=(\eta_1,\eta_2)\in\Xi_4(\bth)\}
\ene
and
\bee\label{tildea}
\tilde a:=\sup\{|\eta_2|,\,\boldeta=(\eta_1,\eta_2)\in\Xi_4(\bth)\}.
\ene
\bel\label{lem:Xi4}
We have: $\tilde p_--p_-=O(\rho^{-1})$ and $\tilde a-a=O(\rho^{-1/3})$.
\enl
\bep
Let $\hat\boldeta:=(p_-,a)$, $\tilde\boldeta:=(\tilde p_-, a)$, and $\check\boldeta:=(p_-,\tilde a)$.
Then all these points belong to  $\overline{\Xi_4(\bth)}$. Thus,
$|\tilde\boldeta|^2-|\hat\boldeta|^2=\tilde p_-^2-p_-^2=O(1)$. Since $p_-\sim\rho$, this implies
$\tilde p_--p_-=O(\rho^{-1})$. Similarly,
$|\check\boldeta|^2-|\hat\boldeta|^2=\tilde a^2-a^2=O(1)$. Since $a\sim\rho^{1/3}$, this implies
$\tilde a-a=O(\rho^{-1/3})$.
\enp
\bel\label{lem:Xi5}
Suppose $R_n\le\rho_n^{1/10}$, $\bxi\in(\Xi_3(\bth)\cup\Xi_4(\bth))$, and let $\bg\in\T'_{15\tilde M}$ be linearly independent of $\bth$. Put $\boldeta:=\bxi+\bg$. Then $||\boldeta|^2-\rho^2|\gg\rho^{4/5}$ and, in particular, $\boldeta\not\in\CA(\rho)$.
\enl
\bep
Since $\bg$ and $\bth$ are linearly independent, $\gamma_1\ne 0$; moreover, corollary \ref{angle1}
implies $|\gamma_1|\gg R_n^{-2}\gg\rho^{-1/5}$. Corollary \ref{cor:Xi3} implies $|\xi_1-\rho|\ll\rho^{-1/3}$.
Thus, $|\eta_1-\rho|\gg\rho^{-1/5}$ and
\bees
|\eta_1^2-\rho^2|\gg\rho^{-1/5}(\eta_1+\rho)\gg\rho^{4/5}.
\enes
Since $\eta_2^2\ll\rho^{2/3}$, this implies $||\boldeta|^2-\rho^2|\gg\rho^{4/5}$, so $\boldeta\not\in\CA(\rho)$.
\enp
Now we make one more definition
\bee\label{Xi5}
\Xi_5(\bth):=\Xi_3(\bth)\setminus(\Xi_4(\bth)+\cup_{j\in\Z}\{j\bth\}).
\ene
\bel\label{lem:Xi6}
Suppose $\bxi\in\Xi_5(\bth)$ and $j\in\Z$. If $\bxi+j\bth\in\L(\bth)$, then $\bxi+j\bth\in\Xi_5(\bth)$.
\enl
\bep
Indeed, our assumptions imply that $\bxi+j\bth\in\Xi_3(\bth)$. Moreover,
\bees
\bxi+j\bth\not\in(\Xi_4(\bth)+\cup_{j\in\Z}\{j\bth\}).
\enes
\enp
Lemmas \ref{lem:Xi5} and \ref{lem:Xi6} immediately imply
\bel\label{lem:Xi7}
Suppose $R_n\le\rho_n^{1/10}$, $\bxi\in\Xi_5(\bth)$, and $\bg\in\T'_{15\tilde M}$.
If $\bxi+\bg\in\CA(\rho)$, then $\bxi+\bg\in\Xi_5(\bth)$.
\enl
\bep
If $\bg$ is linearly independent from $\bth$, this is proved in lemma \ref{lem:Xi5}.
Suppose that $\bg=j\bth$, $j\in\Z$. Then $\bxi+\bg\in\Xi_2(\bth)$, so if we assume
$\bxi+\bg\in\CA(\rho)$, this means that either $\bxi+\bg\in\Xi_3(\bth)$, or $\bxi+\bg\in\Xi_4(\bth)$.
The last possibility contradicts the definition of $\Xi_5(\bth)$. Thus, $\bxi+\bg\in\L(\bth)$ and
now the statement follows from lemma \ref{lem:Xi6}.
\enp
\bel\label{lem:Xi9}
$\Xi_1(\bth)\subset\Xi_5(\bth)$.
\enl
\bep
Definitions of the sets $\Xi_j$ immediately imply that $\Xi_1(\bth)\subset\Xi_3(\bth)$. Thus, it remains
to prove that if $\bxi\in\Xi_4(\bth)$ and $j\in\Z$, $j\ne 0$, we have $\boldeta:=\bxi+j\bth\not\in\CA$.
Without loss of generality we may assume that $\xi_2>0$. Then $\xi_2\in [a,\tilde a]$ (see \eqref{a} and
\eqref{tildea}), and thus lemma \ref{lem:Xi4} implies $\xi_2=a+O(\rho^{-1/3})$. The second condition in
\eqref{conditionsa} implies that the distance between each point of the set $\{a+j|\bth|,j\in\Z,j\ne 0\}$
and $\pm a$ is at least $\frac{|\bth|}{2}$. Thus, $||\eta_2|-|\xi_2||\ge\frac{|\bth|}{3}$ for sufficiently large $\rho$.
Since $\eta_1=\xi_1$, this implies $||\boldeta|^2-|\bxi|^2|\gg \rho^{1/3}|\bth|$. Since $\bxi\in \CA$, this means that
$\boldeta\not\in\CA$. This finishes the proof.
\enp
Now we discuss the relationship between $\Xi_5(\bth_j)$ for various $j$.
\bel\label{lem:Xi8}
Suppose, $R_n\le\rho_n^{1/10}$ and $j_1\ne j_2$. Then $\bigl(\Xi_5(\bth_{j_1})+\T_{15\tilde M}\bigl)\cap \Xi_5(\bth_{j_2})=\emptyset$.
\enl
\bep
Denote $\boldeta_j:=\rho\bn(\bth_{j}^{\perp})$. Then the definition of $\Xi_5$ and corollary \ref{cor:Xi3}
imply that the distance between any point $\bxi\in\Xi_5(\bth_j)$ and $\boldeta_j$ is $O(\rho^{1/3})$.
On the other hand, lemma \ref{angle} implies that $|\boldeta_{j_1}-\boldeta_{j_2}|\gg R_n^{-2}\rho\gg \rho^{1/2}$.
Thus, if   $\bxi_1\in\Xi_5(\bth_{j_1})$ and $\bxi_2\in\Xi_5(\bth_{j_2})$, we have $|\bxi_1-\bxi_2|\gg \rho^{1/2}$.
Since $\rho^{1/2}\gg R_n$, this finishes the proof.
\enp
Now we define
\bee\label{CD}
\CD=\CD(\rho):=\cup_{l=1}^L\Xi_5(\bth_l)
\ene
and
\bee\label{CB}
\CB=\CB(\rho):=\CA(\rho)\setminus\CD(\rho).
\ene
The sets $\Xi_5(\bth)$ are called {\it resonance regions} corresponding to $\bth$. The set
$\CD$ is called the resonance region. Finally, the set $\CB$ is called the non-resonance region.
Obviously, $\CB$ consists of $L$ connected components, each one is located `between' $\bth_l^\perp$ and
$\bth_{l+1}^\perp$ for some $l$, where of course we use the convention that $\bth_{L+1}=\bth_1$.
We call this connected component (located `between' $\bth_l^\perp$ and
$\bth_{l+1}^\perp$) $\CB_l$. More precisely, we define (see figure 4 at the beginning of the section 6)
\bee\label{CBl}
\CB_l:=\{\bx\in\CB,\,\lu\bxi,\bth_l\ru>0\,\&\,\lu\bxi,\bth_{l+1}\ru<0\}.
\ene
We also define
\bee\label{Xi}
\Xi_0(\bth):=\Xi_5(\bth)+\T_{7\tilde M}
\ene
and
\bee
\Xi_0(\CB):=\CB+\T_{\tilde M}.
\ene
\bel\label{lem:Xi10}
We have:
\bee\label{eq1:Xi10}
(\Xi_0(\bth_{j_1})+\T_{\tilde M})\cap \Xi_0(\bth_{j_2})=\emptyset
\ene
when $j_1\ne j_2$ and
\bee\label{eq2:Xi10}
(\Xi_0(\CB)+\T_{\tilde M})\cap \Xi_0(\bth_{j})=\emptyset.
\ene
\enl
\bep
Formula \eqref{eq1:Xi10} follows from Lemma~\ref{lem:Xi8}. Suppose formula \eqref{eq2:Xi10} does not hold. Then there exists
a point $\bxi\in \Xi_5(\bth_j)$ and $\bg\in\T'_{15\tilde M}$ such that $\boldeta:=\bxi+\bg\in\CB\subset\CA$.
Lemma \ref{lem:Xi7} implies that $\boldeta\in\Xi_5(\bth_j)$. This means that $\boldeta\not\in\CB$ in view of
\eqref{CD} and \eqref{CB}.
\enp

Let us introduce more notation. Let
$\CC\subset\Rd$ be a measurable set.
We denote by $\CP^{(\bk)}(\CC)$ the orthogonal projection in $\GH =
L^2([0,2\pi]^d)$ onto the subspace spanned by the exponentials
$\be_{\bxi}(\bx)$, $\bxi\in \CC$, $\{\bxi\}=\bk$.
\bel\label{potential}
For arbitrary set $\CC\subset\Rd$ and arbitrary $\bk$ we have:
\bee\label{eq:potential}
V'\CP^{(\bk)}(\CC)=\CP^{(\bk)}(\CC+\T_1)
V'\CP^{(\bk)}(\CC)
\ene
\enl
\bep This follows from the obvious
observation that if $\bxi=\bm+\bk\in\CC$ and $|\bn|\le R_n$, then
$\bxi+\bn\in\bigl(\CC+\T_1\bigr)$.
\enp
We are going to apply lemma \ref{perturbation2} and now we will
specify what are the projections $P^l_j$. The construction will be
the same for all values of quasi-momenta, so often we will skip
$\bk$ from the superscripts. We denote $P^l:=\CP^{(\bk)}(\Xi_0(\bth_l))$, $l=1,\dots,L$ and
$P^0:=\CP^{(\bk)}(\Xi_0(\CB))$. We also put
\bees
P_j^l:=\CP^{(\bk)}\Bigl((\Xi_5(\bth_l)+\T_{6{\tilde M}+j})\setminus(\Xi_5(\bth_l)+\T_{6{\tilde M}+j-1})\Bigr),\, l=1,\dots,L,\, j=1,\dots,{\tilde M},
\enes
\bees
P_0^l:=\CP^{(\bk)}(\Xi_5(\bth_l)+\T_{6{\tilde M}}),\, l=1,\dots,L,
\enes
\bees
P_j^0:=\CP^{(\bk)}\Bigl((\CB+\T_j)\setminus(\CB+\T_{j-1})\Bigr),\, j=1,\dots,{\tilde M},
\enes
\bees
P_0^0:=\CP^{(\bk)}(\CB).
\enes
Finally, we define $Q:=I-\sum_{l=0}^LP^l$, $\tilde H(\bk):=\sum_{l=0}^LP^lH'(\bk)P^l+QH_0(\bk)Q$, and
$\BJ:=[\la-90v,\la+90v]$.
\bel\label{decomposition}
Let $\mu_n(H'(\bk))\in\BJ$. Then $|\mu_n(\tilde H(\bk))-\mu_n(H'(\bk))|\ll\rho^{-2\tilde M/3}=\rho^{-2M}$.
\enl
\bep
This follows from lemma \ref{perturbation2} and from properties of the sets $\Xi$ formulated in lemmas \ref{lem:Xi1}--\ref{lem:Xi9}.
Indeed, let us check that all the assumptions of lemma \ref{perturbation2} are satisfied. Lemma \ref{lem:Xi10} implies that
if $l\ne n$ then $P^lP^n=P^lV'P^n=0$. The properties $P_j^lV'P_t^l=0$ for $|j-t|>1$ and $P_j^lV'Q=0$ if $j<j_l$ follow from lemma
\ref{potential}. The distance from the spectrum of $QH_0Q$ to $\BJ$ is
greater than $4v$: this follows from the fact that $Q$ is a projection to all the exponentials $\be_{\bxi}$ with $\bxi$ lying
outside of the union $\Xi_0(\CB)\cup\cup_{l}\Xi_0(\bth_l)$ and, thus, satisfying $\bxi\not\in\CA$. Finally, let us show that
the distance from the spectrum of $P_j^l H_0P_j^l$ ($j\ge 1$) to $\BJ$ is greater than $c\rho_n^{1/3}$. When $l=0$, this follows from
Lemma \ref{lem:Xi1} and the fact that $\Lambda(\bth_p)\cap\CB=\emptyset$ for any primitive vector $\bth_p$ from $\T'_{6{\tilde M}}$. Suppose, $l\ne 0$. It is enough to prove that if
\bees
\boldeta\in \Bigl((\Xi_5(\bth_l)+\T_{7\tilde M})\setminus(\Xi_5(\bth_l)+\T_{6\tilde M})\Bigr)
\enes
with $\bth_l\in \T'_{6{\tilde M}}$, then
\bee\label{eq:decomposition}
||\boldeta|^2-\rho^2|\gg\rho^{1/3}.
\ene
Since $\boldeta\in (\Xi_5(\bth_l)+\T_{7\tilde M})$, we can write it as
$\boldeta=\bxi+\bg$ with $\bxi\in\Xi_5(\bth_l)$ and $\bg\in\T_{7\tilde M}$. If $\bg$ and $\bth_l$ are linearly independent,
\eqref{eq:decomposition} follows from lemma \ref{lem:Xi5}. Suppose, $\bg$ is a multiple of $\bth_l$. Let us introduce
coordinates generated by $\bth_l$ as above (after corollary \ref{angle1}). Then we have $\eta_1=\xi_1\in[p_-,p_+]$. Moreover,
since $\boldeta\not\in \Xi_5(\bth_l)+\T_{6\tilde M}$, we have $|\eta_2|\ge a+|\bth_l|$.
Indeed, suppose $|\eta_2|< a+|\bth_l|$. Assume as we can without loss of generality that $\eta_2\ge 0$.
Then $\boldeta-\bth_l\in\Xi_3(\bth_l)$. Since $\bxi\in\Xi_5(\bth_l)$, we have:
$\boldeta-\bth_l=\bxi+j\bth_l\not\in\Xi_4(\bth_l)+\Z\bth_l$. Therefore, $\boldeta-\bth_l\in\Xi_5(\bth_l)$,
so $\boldeta\in\Xi_5(\bth_l)+\T_{6\tilde M}$. This contradiction shows that $|\eta_2|\ge a+|\bth_l|$.

Let us now denote by $\bnu$ the point with coordinates
$\nu_1=p_-$ and $\nu_2=a$. Then $\bnu\in\CA$, so $||\bnu|^2-\rho^2|\ll 1$. But
\bees
|\boldeta|^2-|\bnu|^2\ge \eta_2^2-a^2\ge (a+|\bth_l|)^2-a^2\gg a\ge\rho^{1/3}.
\enes
Thus, $|\boldeta|^2-\rho^2\gg\rho^{1/3}$, which finishes the proof.
\enp
Now we are going to construct mappings $f,g:\R^2\to\R$ with properties stated in the previous section.
Let $\bxi\in\R^2$ with $\{\bxi\}=\bk$. Then we are going to define
\bee\label{f}
f(\bxi)=\mu_p(H(\bk))
\ene
and
\bee\label{g}
g(\bxi)=\mu_p(\tilde H(\bk)),
\ene
where $p=p(\bxi)$ is a natural number chosen in a certain canonical way
so that the mapping $p:\{\bxi\in\R^2,\,\{\bxi\}=\bk\}\to\N$ is a bijection. Leaving aside for a moment the question
of the precise definition of this mapping, we notice that if we define the functions $f$ and $g$ by formulas
\eqref{f} and \eqref{g}, then the properties (i) and (ii) formulated in the previous section will be satisfied
due to lemmas \ref{decomposition} and 
\ref{perturbation2}. So, now we discuss how to define the mapping $p$.
Before doing it, we need more definitions. Let $\bxi\in\CA$. Then $\bxi$ belongs to
exactly one of the sets $\CB$, $\Xi_5(\bth_1)$,..., $\Xi_5(\bth_L)$. If $\bxi\in\CB$, we
define
\bee\label{BUps1}
\BUps(\bxi):=\bxi+\T_{\tilde M}.
\ene
If $\bxi\in\Xi_5(\bth_l)$, we define
\bee\label{BUps2}
\BUps(\bxi)=\BUps(\bxi;\bth_l):=\{\bxi+j\bth_l\in\Xi_3(\bth_l),\,j\in\Z\}+\T_{7\tilde M}.
\ene
We call two vectors $\bxi_1$ and $\bxi_2$ equivalent, if $\BUps(\bxi_1)=\BUps(\bxi_2)$. Note that $\bxi_1,\bxi_2\in\CA$ could be equivalent only if they belong to the same $\Xi_5(\bth_l)$.

Now suppose that $\boldeta\in\R^2$. Then we can define $\BUps(\boldeta)$ in the following way:
if $\boldeta\in\Xi_0(\CB)$, then we have $\boldeta\in\BUps(\bxi)$ for a unique $\bxi\in\CA$ (then $\bxi\in\CB$);
if $\boldeta\in\Xi_0(\bth_l)$, then we have $\boldeta\in\BUps(\bxi)$ for a unique (up to the equivalence) $\bxi\in\CD$ (then $\bxi\in\Xi_5(\bth_l)$). In both these cases we put $\BUps(\boldeta):=\BUps(\bxi)$.
Finally, if $\boldeta\not\in(\Xi_0(\CB)\cup\cup_{l=1}^L\Xi_0(\bth_l))$, we put $\BUps(\boldeta):=\{\boldeta\}$.
We also define $P(\boldeta):=\CP^{(\{\boldeta\})}(\BUps(\boldeta))$.

Lemma \ref{lem:Xi1} implies that the operator $P^0H'(\bk)P^0$ admits a decomposition into invariant subspaces:
\bee\label{BUps3}
P^0H'(\bk)P^0=\bigoplus_{\bxi\in\CB,\{\bxi\}=\bk} P(\bxi)H'(\bk)P(\bxi).
\ene
Similarly, lemma \ref{lem:Xi5} implies that for each $l=1,\dots,L$ we have:
\bee\label{BUps4}
P^lH'(\bk)P^l=\bigoplus P(\bxi)H'(\bk)P(\bxi),
\ene
where the union in the RHS is over all classes of equivalence of $\bxi\in\Xi_5(\bth_l)$ with
$\{\bxi\}=\bk$. Finally, we obviously have:
\bee\label{BUps5}
QH_0(\bk)Q=\bigoplus P(\boldeta)H_0(\bk)P(\boldeta),
\ene
where the union is over all $\boldeta\not\in(\Xi_0(\CB)\cup\cup_{l=1}^L\Xi_0(\bth_l))$, $\{\boldeta\}=\bk$. Moreover,
since all projections $P(\boldeta)$ in \eqref{BUps5} are one-dimensional and we have assumed that
$\int_{\CO}V(\bx)d\bx=0$, we can replace $H_0(\bk)$ with
$H'(\bk)$ for the sake of uniformity so that
\bee\label{BUps5a}
QH_0(\bk)Q=\bigoplus P(\boldeta)H'(\bk)P(\boldeta).
\ene
Thus,
\bee\label{BUps6}
\tilde H(\bk)=\bigoplus P(\boldeta)H'(\bk)P(\boldeta),
\ene
where the union is over all (non-equivalent) $\boldeta\in\R^2$, $\{\boldeta\}=\bk$.

Suppose now $\boldeta\in\R^2$, $\{\boldeta\}=\bk$. Then $|\boldeta|^2$ is an eigenvalue of $P(\boldeta)H_0(\bk)P(\boldeta)$, say
\bee\label{mappingt}
|\boldeta|^2=\mu_t(P(\boldeta)H_0(\bk)P(\boldeta)).
\ene
If $|\boldeta|^2$ is a simple eigenvalue of $P(\boldeta)H_0(\bk)P(\boldeta)$, then this defines the number
$t$ uniquely. Suppose now that $|\boldeta|^2$
is a multiple eigenvalue, say $|\boldeta|^2=|\tilde\boldeta|^2$, $\tilde\boldeta\in\BUps(\boldeta)$, and there are
precisely $t-1$ eigenvalues of $P(\boldeta)H_0(\bk)P(\boldeta)$ below $|\boldeta|^2$.
In this case, we label these eigenvalues according to the crystallographic order of their universal coordinates.
More precisely, we
write $|\boldeta|^2=\mu_t(P(\boldeta)H_0(\bk)P(\boldeta))$ and
$|\tilde\boldeta|^2=\mu_{t+1}(P(\boldeta)H_0(\bk)P(\boldeta))$ if either $\eta_1<\tilde\eta_1$, or
$\eta_1=\tilde\eta_1$ and $\eta_2<\tilde\eta_2$. Thus, we have put into correspondence to any point
$\boldeta$ a number $t=t(\boldeta)$, $t$ varies between $1$ and the number of elements in $\BUps(\boldeta)$.
(Although we will not use this function $t(\boldeta)$ in this section, it will be of much use for us later on).
Next,
we define
\bee\label{mappingnu}
\nu(\boldeta):=\mu_{t(\boldeta)}(P(\boldeta)H'(\bk)P(\boldeta)).
\ene
Due to \eqref{BUps6}, the set
$\{\nu(\boldeta),\,\{\boldeta\}=\bk\}$ coincides with the set of all eigenvalues of $\tilde H(\bk)$
(including multiplicities). Let us label these eigenvalues in an increasing order; in the case of multiple
eigenvalues we, 
as before, label them in accordance with the crystallographic order of their coordinates.
Then to each point $\boldeta$, $\{\boldeta\}=\bk$, we have put into correspondence a number $p=p(\boldeta)$ such that
\bee\label{mappingp}
\nu(\boldeta)=\mu_p(\tilde H(\bk)).
\ene
Thus
defined mapping $p$ is the mapping we are using in the definitions \eqref{f} and \eqref{g}. The rest of this
paper is devoted to introducing the coordinates $(r,\Phi)$ and checking that the conditions of lemma \ref{lem:4}
are satisfied. We start from the non-resonance region $\CB$.

\section {Non-resonance regions}

Suppose that $\bxi\in\CB_l$ (recall that $\CB_l$ is defined in \eqref{CBl} and $\phi(\bx_1,\bx_2)$ is the angle between two non-zero vectors $\bx_1$ and $\bx_2$). Put
$\phi_l:=\frac{\phi(\bth_l,\bth_{l+1})}{2}$. Throughout this section, we fix the coordinate
$(\xi_1,\xi_2)$ introduced after corollary \ref{angle1} and related to $\bth_l$. Namely, we put
$\xi_1=\lu\bxi,\bn(\bth_l^{\perp})\ru$ and
$\xi_2=\lu\bxi,\bn(\bth_l)\ru$.
There is a unique point
$\bnu=\bnu(l)$ satisfying the following two properties: $\lu\bnu,\bn(\bth_l)\ru=a$ and
$\lu\bnu,\bn(\bth_{l+1})\ru=-a$; we have $\nu_1=a\cot\phi_l$, $\nu_2=a$, so $|\bnu|=\frac{a}{\sin\phi_l}$.

\begin{figure}[htb]
    \centering
    \psfrag{L}[][]{$\L(\bth_l)$}
    \psfrag{M}[][]{$\L(\bth_{l+1})$}
    \psfrag{A}[][]{$\CA(\rho)$}
    \psfrag{e}[][]{$\bth_l$}
    \psfrag{k}[][]{$\bth^{\perp}_l$}
    \psfrag{i}[][]{$\bth_{l+1}$}
    \psfrag{j}[][]{$\bth^{\perp}_{l+1}$}
    \psfrag{B}[][]{$\CB_l$}
    \psfrag{g}[][]{$\bnu$}
    \psfrag{F}[][]{$2\phi_l$}
    \includegraphics[width=3in]{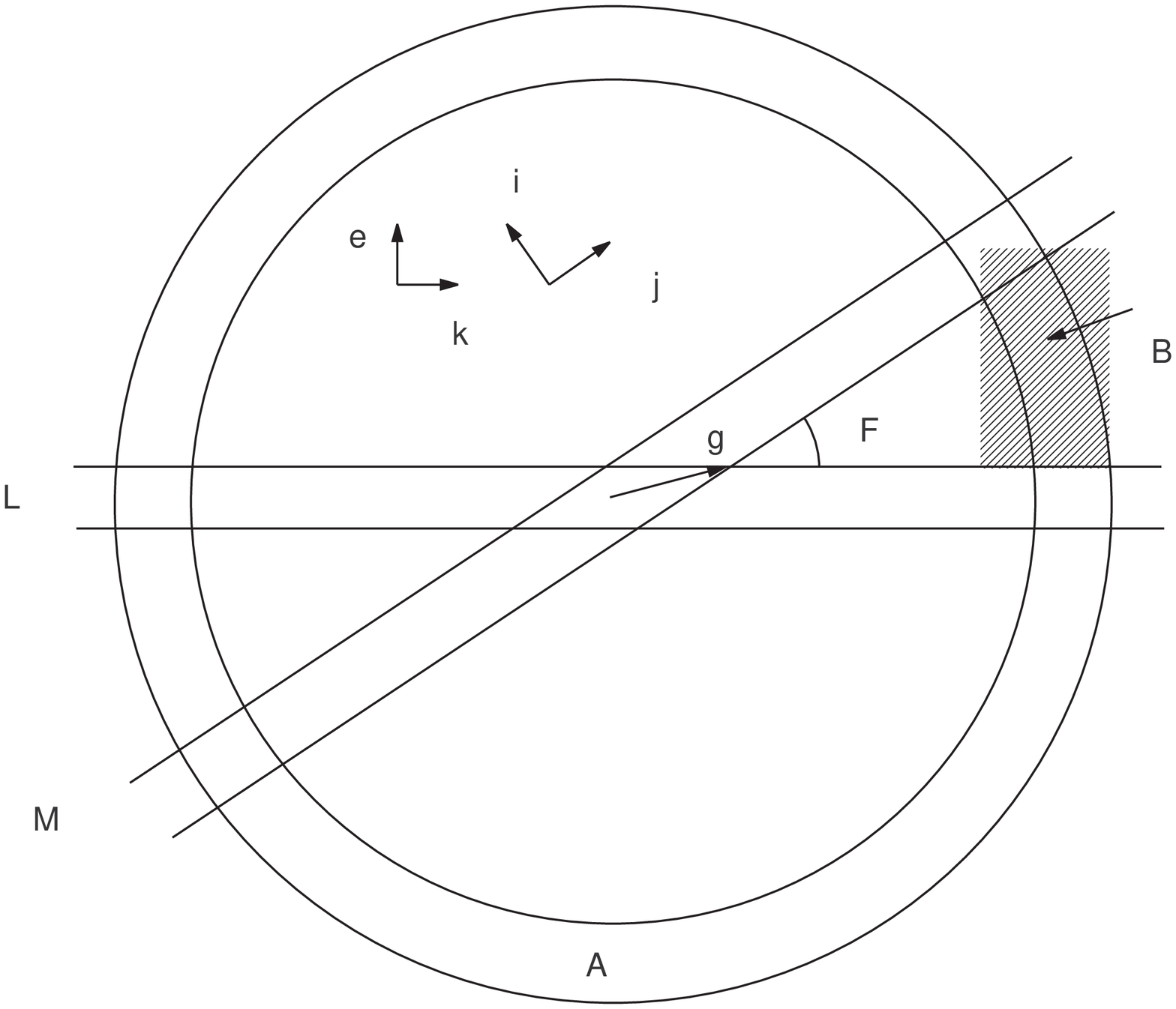}
    \centerline{Figure 4}\label{Fig4}
\end{figure}

We introduce the following pseudo-polar coordinates $(r,\Phi)$ on $\CB_l$:
$r(\bxi):=|\bxi-\bnu_l|$ and $\Phi(\bxi)=
\phi(\bxi-\bnu,\bth_l^\perp)$ when $\bxi\in\CB_l$.
Obviously,
$\Phi(\bxi)\in [0,2\phi_l]=:\CS_l=\CS_l^n$ and $r(\bxi)\sim\rho$ when $\bxi\in\CB_l$.
We also have the following formulas:
$\xi_1=\nu_1+r(\bxi)\cos(\Phi(\bxi))$ and $\xi_2=\nu_2+r(\bxi)\sin(\Phi(\bxi))$.
Therefore,
\bee\label{polar1}
\bes
|\bxi|^2
&
=(r(\bxi)\cos(\Phi(\bxi))+|\bnu|\cos\phi_l)^2+
(r(\bxi)\sin(\Phi(\bxi))+|\bnu|\sin\phi_l)^2
\\
&
=(r(\bxi)\cos(\Phi(\bxi))+a\frac{\cos\phi_l}{\sin\phi_l})^2+
(r(\bxi)\sin(\Phi(\bxi))+a)^2.
\end{split}
\ene
This implies that there is a complete asymptotic formula:
\bee\label{polar3}
|\bxi|\sim r(\bxi)\left(1+\sum_{j=1}^{\infty} \tilde
b_j(\Phi(\bxi))r(\bxi)^{-j}\right)
\ene
with $\tilde b_j=\tilde b_j(\Phi(\bxi))\ll a^{j}R_n^{2j}\ll\rho_n^{\frac{j}{2}}$ as $r(\bxi)\to\infty$, uniformly over $\bxi\in\CB_l$,
and this formula can be differentiated once. (Here we assumed that $R_n\ll\rho_n^{1/12}$.)

The following lemma was proved in \cite{Par} (Lemma 6.1 there):

\bel\label{eigenvalues1} Let $R_n\ll\rho_n^{1/24}$.
Then the following asymptotic formula holds:
\bee\label{eq:eigenvalues1}
\bes
&g(\bxi)\sim|\bxi|^2\\
&+\sum_{s=1}^{\infty} \sum_{\boldeta_1,\dots,\boldeta_s\in \T'_{\tilde M}}
\sum_{m_1+\dots+m_s\ge 2}A_{m_1,\dots,m_s}
\lu\bxi,\boldeta_1\ru^{-m_1}\dots\lu\bxi,\boldeta_{s}\ru^{-m_s}
\end{split}
\ene
in a sense that for each natural $K$ we have
\bee\label{eq:eigenvalues1bis}
\bes
&g(\bxi)=|\bxi|^2\\
&+\sum_{s=1}^{3K} \sum_{\boldeta_1,\dots,\boldeta_s\in \T'_{\tilde M}}
\sum_{m_1+\dots+m_s\ge 2}A_{m_1,\dots,m_s}
\lu\bxi,\boldeta_1\ru^{-m_1}\dots\lu\bxi,\boldeta_{s}\ru^{-m_s}+o(\rho^{-K})
\end{split}
\ene
uniformly over $R_n\ll\rho_n^{1/24}$ and $\bxi\in\CB$.
Here, $A_{m_1,\dots,m_p}$ is a polynomial of the Fourier
coefficients $\hat V(\boldeta_j)$ and $\hat
V(\boldeta_j-\boldeta_l)$ of the potential and the exponents $m_1,\dots,m_s$ are positive integers.
Moreover,
\bee\label{Aestimates}
|A_{m_1,\dots,m_p}|\ll 1
\ene
uniformly over $n$ (but with the implied constant depending on $V$ and $m_1$,...,$m_p$).
\enl
\ber
Estimate \eqref{Aestimates} was not stated in \cite{Par}, but it follows easily from the proof of Lemma 6.1 there.
\enr
\bec
We have:
\bee\label{eq:neweigenvalues1}
g(\bxi)=|\bxi|^2
+G(\bxi)+o(\rho^{- M}),
\ene
where
\bee\label{eq:neweigenvalues2}
G(\bxi):=\sum_{s=1}^{\tilde M} \sum_{\boldeta_1,\dots,\boldeta_s\in \T'_{\tilde M}}
\sum_{2\le m_1+\dots+m_s\le \tilde M}A_{m_1,\dots,m_s}
\lu\bxi,\boldeta_1\ru^{-m_1}\dots\lu\bxi,\boldeta_{s}\ru^{-m_s}.
\ene
\enc
\ber
Since, as we have seen in lemma \ref{lem:2} and corollary \ref{cor:3}, the terms of
order $O(\rho^{-M})$ do not contribute to asymptotic formula \eqref{eq:main_lem1}, we
can re-define
\bee\label{eq:newg}
g(\bxi):=|\bxi|^2
+G(\bxi).
\ene
\enr

\bel\label{eq:eigenvalues2} Assume $R_n\ll\rho_n^{1/20}$.
For each $m\in\N$ and $\boldeta\in\T'_{6\tilde M}$ such that $\boldeta$ is not a multiple of $\bth_l$ or
$\bth_{l+1}$ there is a complete asymptotic formula:
\bee\label{eq:eigenvalues21}
\lu\bxi,\boldeta\ru^{-m}\sim\sum_{j=m}^\infty r(\bxi)^{-j} c_j^m(\Phi(\bxi))
\ene
uniformly over $\bxi\in\CB_l$, where $|c_j^m|\ll\rho_n^{j/2}$. 
Similar formulas are valid if $\boldeta$ is a multiple of $\bth_{l+1}$ and $0\le\Phi(\bxi)\le\phi_l$,
or if $\boldeta$ is a multiple of $\bth_{l}$ and $\phi_l\le\Phi(\bxi)\le 2\phi_l$.
\enl
\bep
We have: $\bxi=\bnu+r(\bxi)\bn(\bxi-\bnu)$.
Therefore, $\lu\bxi,\boldeta\ru=\lu\bnu,\boldeta\ru+r(\bxi)\lu\bn(\bxi-\bnu),\boldeta\ru$.
Our constructions and corollary \ref{angle1} imply 
$|\lu\bn(\bxi-\bnu),\boldeta\ru|\gg R_n^{-2}$. Thus,
\bee\label{eq:eigenvalues22}
\bes
\lu\bxi,\boldeta\ru^{-1}&=r^{-1}\lu\bn(\bxi-\bnu),\boldeta\ru^{-1}(1+r^{-1}\lu\bn(\bxi-\bnu),\boldeta\ru^{-1}\lu\bnu,\boldeta\ru)^{-1}\\
&=r^{-1}\lu\bn(\bxi-\bnu),\boldeta\ru^{-1}\sum_{j=0}^\infty
(-1)^j r^{-j}\lu\bn(\bxi-\bnu),\boldeta\ru^{-j}\lu\bnu,\boldeta\ru^j
\end{split}
\ene
and $\lu\bn(\bxi-\bnu),\boldeta\ru^{-j}\lu\bnu,\boldeta\ru^j\ll R_n^{3j}\rho_n^{j/3}\ll\rho_n^{j/2}$. Now
formula \eqref{eq:eigenvalues21} is obtained from \eqref{eq:eigenvalues22} by raising both sides
to the $m$-th power. The proof of the last two statements is similar.
\enp

Unfortunately, lemma \ref{eq:eigenvalues2} does no longer hold if $\boldeta$ is a multiple of $\bth_l$ or $\bth_{l+1}$ and $\Phi(\bxi)$ is close to $0$ or $2\phi_l$ respectively. Therefore, we cannot apply lemma \ref{lem:4} without modifications. This means, we need to do some extra work. We can assume, without loss of generality, that
$\phi_l\le 1/100$, which is certainly the case for sufficiently large $n$.

Let us fix an angle $\Phi$ for a moment, $0\le\Phi\le 2\phi_l$ and let $\bxi=(r,\Phi)$
where only $r$ varies. Denote by $r_0=r_0(\Phi;\rho)$ a unique value of $r$ which corresponds to $\bxi$ satisfying $|\bxi|^2=\rho^2$. It is easy to check  
that the partial
derivative $\frac{\partial G}{\partial r}=O(\rho^{-4/3})$. Therefore, there is a unique value of $r$
such that corresponding point $\bxi=(r,\Phi)$ satisfies $g(\bxi)=\rho^2$; we denote this
value of $r$ by $r_1=r_1(\Phi;\rho)$.

\bel\label{lem:next1}
There is an asymptotic decomposition
\bee\label{eq:next1}
r_0\sim \rho(1+\sum_{j=1}^\infty p_j\rho^{-j}),
\ene
where $p_j=p_j(\Phi)=O(\rho_n^{j/2})$ uniformly over $\Phi$.
Moreover, we have $p_1=-a\cos(\phi_l-\Phi)(\sin\phi_l)^{-1}$,
$p_{2m}={1/2\choose m}(-1)^m a^{2m}\sin^{2m}(\phi_l-\Phi)(\sin\phi_l)^{-2m}$, and $p_{2m+1}=0$
for $m\in\N$.
\enl
\bep
This follows from the explicit formula which can be easily obtained using the cosine theorem:
\bee\label{eq:next2}
r_0=-a\cos(\phi_l-\Phi)(\sin\phi_l)^{-1}+\sqrt{\rho^2-a^2\sin^2(\phi_l-\Phi)(\sin\phi_l)^{-2}}.
\ene
\enp
\bel\label{lem:next2}
We have:
\bee\label{eq:next3}
\vol(\hat A^+\cap\CB_l)-\vol(\hat A^-\cap\CB_l)=\frac{1}{2}\int_0^{2\phi_l}
(r_1(\Phi;\rho)^2- r_0(\Phi;\rho)^2)d\Phi.
\ene
\enl
\bep
Integrating in polar coordinates, we have:
\bee\label{eq:next4}
\vol(\hat A^+\cap\CB_l)-\vol(\hat A^-\cap\CB_l)=
\int_0^{2\phi_l}d\Phi\int_0^{r_1}rdr-\int_0^{2\phi_l}d\Phi\int_0^{r_0}rdr
=
\int_0^{2\phi_l}\frac{r_1^2-r_0^2}{2}d\Phi.
\ene
\enp
The last two lemmas show that in order to compute $\vol(\hat A^+\cap\CB_l)-\vol(\hat A^-\cap\CB_l)$, it remains to compute $r_1$. We do it using the sequence of approximations.
Assume as above that $\Phi$ is fixed.
Put $\tilde r_0:=r_0$ and $\bxi_0:=(\tilde r_0,\Phi)$. The further elements of the sequence are defined like this: $\bxi_{m+1}=(\tilde r_{m+1},\Phi)$ is a unique point
satisfying $|\bxi_{m+1}|^2=\rho^2-G(\bxi_m)$.
\bel\label{lem:next3}
For each $m\in\N$ we have:
\bee\label{eq:next5}
r_1=\tilde r_m+O(\rho^{-m}).
\ene
\enl
\bep
Put
\bee
H(r):=-a\cos(\phi_l-\Phi)(\sin\phi_l)^{-1}
+\sqrt{\rho^2-G(r,\Phi)-a^2\sin^2(\phi_l-\Phi)(\sin\phi_l)^{-2}}.
\ene
Then $H'(r)=O(\rho^{-4/3})$. Moreover, $r_1$ is a unique solution of equation $r_1=H(r_1)$. Thus, Banach contraction mapping theorem tells us
that the sequence $\tilde r_m$ satisfying $\tilde r_{m+1}=H(\tilde r_m)$ converges to
$r_1$ and $|r_1-\tilde r_{m+1}|\ll\rho^{-1}|r_1-\tilde r_m|$. Since $r_1=r_0+O(1)$,
this finishes the proof.
\enp
\bec\label{cor:next1}
We have:
\bee\label{eq:next6}
\vol(\hat A^+\cap\CB_l)-\vol(\hat A^-\cap\CB_l)=\frac{1}{2}\int_0^{2\phi_l}
(\tilde r_{M+1}(\Phi;\rho)^2- r_0(\Phi;\rho)^2)d\Phi+O(\rho^{-M}).
\ene
\enc
Analogously to \eqref{eq:next2}, we have:
\bee\label{eq:next7}
\tilde r_{m+1}=-a\cos(\phi_l-\Phi)(\sin\phi_l)^{-1}
+\sqrt{\rho^2-G(r_m,\Phi)-a^2\sin^2(\phi_l-\Phi)(\sin\phi_l)^{-2}}.
\ene
Taking into account \eqref{eq:neweigenvalues2}, \eqref{eq:eigenvalues22}, \eqref{eq:next2}, \eqref{eq:next7}, and lemma~\ref{eq:eigenvalues2}, we obtain that
\begin{equation}\label{below}
\bes
\tilde{r}_{M+1}&=r_0+\rho^{-1}\sum\limits_{j,s\geq0;j+s\geq2}C_{j,s,M}(\Phi)\rho^{-j}\langle\bxi_0,\bn(\bth_l)\rangle^{-s}\\
&=r_0+\rho^{-1}\sum\limits_{j,s\geq0;2\le j+s\leq2\tilde M}C_{j,s,M}(\Phi)\rho^{-j}\langle\bxi_0,\bn(\bth_l)\rangle^{-s}+O(\rho^{-M-1})
\end{split}
\end{equation}
for $0\leq\Phi\leq\phi_l$.
Similarly,
\begin{equation}\label{above}
\bes
\tilde{r}_{M+1}&=r_0+\rho^{-1}\sum\limits_{j,s\geq0;j+s\geq2}\tilde{C}_{j,s,M}(\Phi)\rho^{-j}\langle\bxi_0,\bn(\bth_{l+1})\rangle^{-s}\\
&=r_0+\rho^{-1}\sum\limits_{j,s\geq0;2\le j+s\leq2\tilde M}\tilde C_{j,s,M}(\Phi)\rho^{-j}\langle\bxi_0,\bn(\bth_l)\rangle^{-s}+O(\rho^{-M-1})
\end{split}
\end{equation}
for $\phi_l\leq\Phi\leq2\phi_l$.
Here, $C_{j,s,M}(\Phi)$ and $\tilde{C}_{j,s,M}(\Phi)$ are polynomials of
$\cos(\varphi_l - \Phi)$, $\sin(\varphi_l - \Phi)$, and expressions of the form
$(\langle \bn(\bxi - \bnu),
\bn(\boldeta_t)\rangle)^{-
1}$, where ${\boldeta}_t\in \Theta'_{\tilde{M}}$ are not multiples of
 $\bth_l$
 or $\bth_{l+1}$. Each
term in the polynomial $C_{j,s,M}(\Phi)$ and $\tilde{C}_{j,s,M}(\Phi)$ is $O(\rho_n^{j/2})$, and the number of such
terms is $O(R_n^{2(j+s)})$.

Next, we note that $\langle\bxi_0,\bn(\bth_l)\rangle=a+r_0\sin\Phi$ and $\langle\bxi_0,\bn(\bth_{l+1})\rangle=-a-r_0\sin(2\phi_l-\Phi)$.
Thus, in order to use Corollary \ref{cor:next1}, we need to compute integrals of the form
\bee\label{Cterms}
\int\limits_0^{\varphi_l} \frac{C_{j,s,M}(\Phi)r_0(\Phi)^\delta\,d\Phi}{(a+r_0
(\Phi)\sin(\Phi))^s},\ \ \
\int\limits_{\varphi_l}^{2\varphi_l} \frac{\tilde{C}_{j,s,M}(\Phi)r_0(\Phi)
^\delta\,d\Phi}{(-a-r_0(\Phi)\sin(2\varphi_l-\Phi))^s},\ \ \delta=0,1.
\ene
Taking into account Lemma \ref{lem:next1} and properties of $C_{j,s,M}(\Phi),\ \tilde{C}_{j,s,M}(\Phi)$
stated above, we can decompose all functions of $\Phi$ in \eqref{Cterms} into Taylor's series in
the neighborhoods of $\Phi=0$ and $\Phi=2\varphi_l)$. At the same time we apply the following transform of the denominator:
\bee\label{transform}
\begin{split}&
(a+r_0(\Phi)\sin(\Phi))^{-s}=(a+\rho\Phi(1+p(\rho,\Phi))(1+\phi(\Phi)))^{-s}=\cr &
(a+\rho\Phi)^{-s}\left(1+\frac{\Phi}{a/\rho+\Phi}(p(\rho,\Phi)+\phi(\Phi)+p(\rho,\Phi)\phi(\Phi))\right)^{-s}=\cr &
\rho^{-s}(a/\rho+\Phi)^{-s}\left(1+\sum\limits_{k=1}^{\infty}(-1)^k
\left(\frac{\Phi}{a/\rho+\Phi}(p(\rho,\Phi)+\phi(\Phi)+p(\rho,\Phi)\phi(\Phi))\right)^k\right)^{s},
\end{split}
\ene
where (see Lemma~\ref{lem:next1})
$$
p(\rho,\Phi):=\sum_{k=1}^\infty p_k(\Phi)\rho^{-k}
$$
and
$$
\phi(\Phi):=\sum\limits_{k=1}^{\infty}\frac{(-1)^k}{(2k+1)!}\Phi^{2k};
$$
recall that we are assuming that $\phi_l\le 1/100$, so that there is no doubt about the convergence of the last series
in \eqref{transform}.
Thus, decomposing $p_k(\Phi)$ into Taylor's series we reduce the problem to computing the following model integrals:
$$
\int\limits_0^{\varphi_l}\frac{\Phi^k\,d\Phi}{(a/\rho + \Phi)^m}.
$$
After substitution $x:=a/\rho + \Phi$ we can explicitly calculate these integrals. Note, that if $1\leq m\leq k+1$ then the term $\ln\rho$ appears. Combining together all contributions we obtain the following lemma, which is the main result of this section:

\bel\label{nonresonance} Assume that $R_n\ll\rho_n^{1/24}$. Then
\bee\label{eq:nextlast}
\vol(\hat A^+\cap\CB_l)-\vol(\hat A^-\cap\CB_l)=\sum\limits_{j=1}^{\tilde{M}}C_j\rho^{-j}+\ln\rho\,\sum\limits_{j=2}^{\tilde{M}}\tilde{C}_j\rho^{-j}+O(\rho^{-M}),
\ene
where $|C_1|\ll\rho_n^{-1/6}$, $|C_j|,\,|\tilde{C}_j|\ll\rho_n^{2j/3},\ j\geq2$.
\enl

\section{Resonance regions}

We now consider $\boldeta\in\CD$ 
and try to compute $g(\boldeta)$.
The key result in this section is corollary \ref{cor:7.12}, where we compute $g(\boldeta)$ in this setting.
In the rest
of this section, we fix $l$ and omit it from the notation, so that $\bth:=\bth_l$. We also assume that $n$ is fixed and
will frequently omit it from the notation.
As above, we
introduce the coordinates $\boldeta=(\eta_1,\eta_2)$ so that $\eta_1=\lu \boldeta,\bn(\bth^\perp)\ru $ and $\eta_2=\lu\boldeta,\bn(\bth)\ru$.
We obviously have $\eta_1\sim\rho$ and $|\eta_2|\ll\rho^{1/3}$. It is convenient to denote $r:=\eta_1$ and $\Phi:=\eta_2$, to indicate that
$(r,\Phi)$ are going to play the same role as in corollary \ref{cor:3} (or rather remark \ref{rem:alternative}) and lemma \ref{lem:4}; note that
$(r,\Phi)$ satisfy all properties of remark \ref{rem:alternative}. We also fix an element $\bxi\in\Xi_5(\bth_l)$ in each set $\BUps$ and assume that
$\boldeta\in\BUps(\bxi)$; 
the point is that we will frequently treat $\bxi$ as fixed and study how $g(\boldeta)$ varies when $\boldeta$
runs over $\BUps(\bxi)$.

Let
$\bnu_0=0,\bnu_1,\dots,\bnu_p$
be a complete system
of representatives of $\T_{7\tilde M}$ modulo $\bth$. That means that $\bnu_j\in\T_{7\tilde M}$ and
each vector $\bg\in\T_{7\tilde M}$ has
a unique representation $\bg=\bnu_j+ m\bth$, $m\in\Z$. We denote the coordinates of $\bnu_j$ by $(\nu'_j,\nu''_j)$
and put $\BPsi_j=\BPsi_j(\bxi):=\bigl(\bxi+\bnu_j+(\Z\bth)\bigr)\cap\BUps(\bxi)$.
Then each set $\BPsi_j$ consists of points having the same first coordinate; the distances between points in $\BPsi_j$ are multiples
of $|\bth|$. Moreover,
\bee
\BUps(\bxi)=\bigcup_j \BPsi_j,
\ene
and this is a disjoint union.

Let us compute diagonal elements of $H(\bxi):=P(\bxi)H'(\bk)P(\bxi)$, where $\bk=\{\bxi\}$. Put 
$\GH(\bxi):=P(\bxi)\GH$, so that $H(\bxi)$ can be thought of as an operator acting in $\GH(\bxi)$.

Let $\boldeta\in\BUps(\bxi)$.
Then $\boldeta$ can be uniquely decomposed as
\bee\label{etadecomposition}
\boldeta=\bxi+m\bth+\bnu_j
\ene
with $m\in\Z$. 
Recall that $H(\bxi)=P(\bxi)(H_0(\bk)+V')P(\bxi)$ and
$H_0(\bxi)\be_{\boldeta}=|\boldeta|^2\be_{\boldeta}$ whenever $\boldeta\in\BUps(\bxi)$. 
We obviously have:
\bee\label{AB}
\bes
|\boldeta|^2&=
|\bxi+\bnu_j+m\bth|^2=
(r+\nu'_j)^2+(\xi_2+\nu''_j+m|\bth|)^2\\
&=r^2+
2\nu'_j r+
{\nu'_j}^2+(\xi_2+\nu''_j+m|\bth|)^2.
\end{split}
\ene
This simple computation
implies that
\bee\label{newpencil}
H(\bxi)=r^2I+rA+B.
\ene
Here, $A=A(\bxi)$ and
$B=B(\bxi)$ are self-adjoint operators acting in $P(\bxi)\GH$
 in the following way:
\bee\label{operatorA0}
A=2\sum_{j=0}^p\nu'_j\CP^{(\bk)}(\BPsi_j);
\ene
in other words, for $\boldeta\in\BPsi_j$ we have
\bee\label{operatorA}
A\be_{\boldeta}=2\nu'_j \be_{\boldeta}
=2(\boldeta-\bxi)_1 \be_{\boldeta},
\ene
and
\bee\label{operatorB}
B\be_{\boldeta}=({\nu'_j}^2+(\xi_2+\nu''_j+m|\bth|)^2+P(\bxi)V')\be_{\boldeta}
\ene
for all $\boldeta\in \BPsi_j(\bxi)$ with $\bnu_j$ and $m$ being defined by
\eqref{etadecomposition}.
These definitions imply that
\bee\label{GV}
\GV:=\ker
A=\CP^{(\bk)}(\BPsi_0)\GH(\bxi).
\ene
Notice that
\bee\label{newboundA}
\|A(\bxi)\|\ll R_n\ll\rho_n^{1/3}
\ene
and
\bee\label{newboundB}
\|B(\bxi)\|\ll \rho_n^{2/3}.
\ene
Let us state more properties of $A$ and $B$.
\bel\label{lem:A}
Let $\mu$ be a non-zero eigenvalue of $A$. Then $|\mu|\gg R_n^{-2}$.
\enl
\bep
Formula \eqref{operatorA} implies that the eigenvalues of $A$ equal $\{\nu'_j\}$. We also have:
$\nu'_j=\lu\bnu_j,\bn(\bth^\perp)\ru$. Now the statement follows from corollary \ref{angle1}.
\enp
Let us define $\tilde P$ to be the orthogonal projection onto $\GV=\ker A$ acting in $\GH(\bxi)$
and $\tilde B:=\tilde P B\tilde P:\GV\to\GV$.
Note that considering operators acting in $\GV$ means considering only $j=0$ (and thus $\bnu_0=0$)
in \eqref{operatorA0} and \eqref{operatorB}. Thus, in particular, we have:
\bee\label{operatortildeB}
{\tilde B}\be_{\boldeta}=((\xi_2+m|\bth|)^2+\tilde P(\bxi)V')\be_{\boldeta}
\ene
if $\boldeta=\bxi+m\bth\in\Psi_0(\bxi)$.
We also denote $\hat n_2:=[\frac{\xi_2}{|\bth|}]$
and $\hat k_2:=\{\frac{\xi_2}{|\bth|}\}$ (note that $\hat n_2$ is not the second coordinate of $\bn=[\bxi]$; this is why
we did not call it $n_2$).
\bel\label{lem:tildeB}
We have:
\bee\label{eq:tildeB}
\mu_{j+2}(\tilde B)-\mu_j(\tilde B)\gg 1
\ene
uniformly over $j$, $n$, $l$ and $\bxi\in\Xi_5(\bth)$.
\enl
\bep
Denote by $T$ the number of elements in $\{\bxi+j\bth,\,j\in\Z\}\cap\Lambda(\bth)$.
Inequality \eqref{eq:tildeB} obviously holds if $j\ge T-2$. Indeed, denote by $\tilde B_0$ the operator
$\tilde B$ with potential $V$ being identical zero. Then we have $|\mu_{j}(\tilde B)-\mu_j(\tilde B_0)|\le v$.
On the other hand, it is easy to check that
$\mu_{j+2}(\tilde B_0)-\mu_j(\tilde B_0)\gg a\sim\rho^{1/3}$.

Let us assume now that $j<T-2$. Then we will compare eigenvalues of operator $\tilde B$ with the eigenvalues of
a certain one-dimensional Sturm-Liouville operator.
Let $\hat\BUps(\bxi):=\{\bxi+j\bth,\,j\in\Z\}$, $\hat P(\bxi):=\CP^{(\bk)}(\hat\BUps(\bxi))$,
and $\hat\GH(\bxi):=\hat P(\bxi)\GH$.
Consider an operator $\hat B=\hat B_n$ (later on in the proof, we will need to remember that these operators depend on $n$)
acting in $\hat\GH(\bxi)$ by the formula
\bee\label{operatorhatB}
{\hat B}\be_{\boldeta}=((\xi_2+j|\bth|)^2+\hat P(\bxi)V')\be_{\boldeta}
\ene
for each $\boldeta=\bxi+j\bth\in\hat\BUps(\bxi)$. Then, in the same way as we proved lemma \ref{decomposition} using lemma
\ref{perturbation2}, we can show that if $j\le T$, we have
\bee\label{eq:tildeB1}
|\mu_j(\tilde B)-\mu_j(\hat B)|\ll\rho_n^{-(\tilde M-1)/3}.
\ene
However, the operator $\hat B$ is unitary equivalent to a one-dimensional Schr\"odinger operator $-y''+{\tilde V}$
on the interval $[0,2\pi|\bth|^{-1}]$ with a potential
\bee\label{W}
{\tilde V}={\tilde V}_{\bth,R_n}=\sum_{m\in\Z,\ |m\bth|\le R_n}\left(\frac{|\bth|}{2\pi}\right)^{1/2}e^{ix(\xi_2+m|\bth|)}\hat V(m\bth)
\ene
and quasi-periodic
boundary conditions  $y(\frac{2\pi}{|\bth|})=e^{2\pi \hat k_2i}y(0)$ and $y'(\frac{2\pi}{|\bth|})=e^{2\pi \hat k_2i}y'(0)$.
Indeed, the isometry  $S$ which establishes this unitary equivalence is given by
$S:\  \be_{\bxi+m\bth}\mapsto \left(\frac{|\bth|}{2\pi}\right)^{1/2}e^{ix(\xi_2+m|\bth|)}$. Standard results
about one-dimensional Schr\"odinger operators (see e.g. \cite{ReeSim}) imply that
\bee\label{eq:hatB}
\mu_{j+2}(\hat B)-\mu_j(\hat B)\gg |\bth|^{2}.
\ene
The simplest way to see why this inequality holds is to notice that the distance between eigenvalues of $\hat B$ and
the unperturbed eigenvalues $\{(m+{\hat k}_2)^2|\bth|^2\}_{m\in\Z}$ is at most the
$L_\infty$-norm of the potential ${\tilde V}_{\bth,R_n}$.
This shows that \eqref{eq:hatB} holds when $j\ge C v$, whereas for finitely many $j$ satisfying
$j<C v$ we can use the fact that $\mu_{j+2}(\hat B)\ne \mu_j(\hat B)$, since an eigenvalue of a one-dimensional differential operator of second order cannot have multiplicity three.
Inequalities \eqref{eq:hatB} and \eqref{eq:tildeB1} prove \eqref{eq:tildeB} for $j\le T$. Let us prove that this estimate is
uniform in $j$, $n$, and $l$. Indeed, the uniformity of \eqref{eq:tildeB1} follows from lemma
\ref{perturbation2}. Consider \eqref{eq:hatB}. Uniformity in $j$ follows from the remark after \eqref{eq:hatB}.
It follows immediately from the same remark that \eqref{eq:hatB} is uniform when $L_\infty$-norm of the potential ${\tilde V}_{\bth,R_n}$ satisfies
\bee\label{W1}
||{\tilde V}_{\bth,R_n}||_\infty\le \frac{|\bth|^{2}}{8}.
\ene
Since the potential $V$ is infinitely smooth, we have $|\hat V(\bg)|\ll |\bg|^{-2}$, which shows that there are
only finitely many $\bth$ for which \eqref{W1} is not satisfied. This shows uniformity of  \eqref{eq:hatB} in $l$. It remains to
prove the uniformity of \eqref{eq:hatB} in $n$ when $\bth$ is fixed. First, we notice that \eqref{eq:hatB} holds for sufficiently
large $j\ge j_0$, where $j_0$ depends only on $||V||_\infty$, but not on $n$.
Suppose now that \eqref{eq:hatB} is not uniform in $n$. Then there is a value of $j$ such that 
\bee\label{limits}
\lim_{n\to\infty}\mu_{j+2}(\hat B_n)=\lim_{n\to\infty}\mu_{j+1}(\hat B_n)=\lim_{n\to\infty}\mu_{j}(\hat B_n)=:\mu
\ene
(strictly speaking, we need to pass to a subsequence $n_k$ if necessary). However, these limits are the eigenvalues of the limit
operator $\hat B_\infty$ with the potential
\bee\label{Winfty}
{\tilde V}_{\bth,\infty}=\sum_{m\in\Z}\left(\frac{|\bth|}{2\pi}\right)^{1/2}e^{ix(\xi_2+m|\bth|)}\hat V(m\bth).
\ene
The required result now follows from the fact we already used above that a second order one-dimensional differential operator $\hat B_\infty$ cannot
have an eigenvalue of multiplicity three.
\enp

Our next task is to compare eigenvalues of $H(\bxi)$ and $H(\bxi')$ when $\bxi$ and $\bxi'$ are two different vectors lying in
$\Xi_5(\bth)$. This is not a straightforward task, since these operators act in different Hilbert spaces ($\GH(\bxi)$ and
$\GH(\bxi')$ correspondingly). Thus, first of all we need to be able to map these Hilbert spaces onto each other. The natural
candidate for such a mapping is
\bee\label{mappingF}
F_{\bxi,\bxi'}(\be_{\boldeta})=\be_{\boldeta+\bxi'-\bxi}.
\ene
Ideally, we would like this mapping to act as follows: $F_{\bxi,\bxi'}: \GH(\bxi)\to \GH(\bxi')$ and be an isomorphism.
Unfortunately, in general this is not the case since the sets $\BUps(\bxi)$ and $\BUps(\bxi')$ can contain different number
of elements. In fact, it may well happen that $\boldeta\in\BUps(\bxi)$, but $(\boldeta+\bxi'-\bxi)\not\in\BUps(\bxi')$. However,
the mapping $F$ has the suggested property in one very important special case: when $\Phi(\bxi)=\Phi(\bxi')$ (in other words, when
the second coordinates of $\bxi$ and $\bxi'$ coincide). Indeed, suppose that $\Phi(\bxi)=\Phi(\bxi')$. Then obviously
\bees
\{\bxi+j\bth_l\in\Xi_3(\bth_l),\,j\in\Z\}+(\bxi'-\bxi)=\{\bxi'+j\bth_l\in\Xi_3(\bth_l),\,j\in\Z\}.
\enes
Thus, we also have
\bees
\BUps(\bxi)+(\bxi'-\bxi)=\BUps(\bxi'),
\enes
and so the mapping $F_{\bxi,\bxi'}$ is an isometry between $\GH(\bxi)$ and $\GH(\bxi')$ with $F_{\bxi,\bxi'}^{-1}=F_{\bxi',\bxi}$. Moreover, if we look carefully on
formulas \eqref{operatorA} (the first equality there) and \eqref{operatorB}, we realize that the definitions of operators $A(\bxi)$ and $B(\bxi)$
do not depend on $\xi_1$, so we have $F_{\bxi',\bxi}A(\bxi')F_{\bxi,\bxi'}=A(\bxi)$ and, similarly,
$F_{\bxi',\bxi}B(\bxi')F_{\bxi,\bxi'}=B(\bxi)$. Thus, all operators $A(\bxi)$
are unitary equivalent when $\bxi$ runs
along any horizontal line $\Phi(\bxi)=\Phi_0$; the same statement holds for $B(\bxi)$. It is convenient to think of all such operators as being identical operators
$A(\Phi_0)$ and $B(\Phi_0)$ acting in the same Hilbert space $\GH(\Phi_0)$. We also notice that if $\Phi(\bxi)=\Phi(\bxi')$, then
the isometry $F_{\bxi',\bxi}$ leaves the function $t$ (defined after \eqref{mappingt}) invariant. This means that whenever
$\boldeta\in\BUps(\bxi)$ and $\boldeta'=\boldeta+\bxi'-\bxi\in\BUps(\bxi')$, we have $t(\boldeta)=t(\boldeta')$;
this is true even if $|\boldeta|^2$ is a multiple eigenvalue of $H_0(\boldeta)$ (and, correspondingly,
 $|\boldeta'|^2$ is a multiple eigenvalue of $H_0(\boldeta')$).

Denote $\CS=\CS^{l+L}_n:=[-a,a]$, $l=1,\dots,L$ (recall that $\CS^l$ were already introduced in the previous section).
Now it seems to be a straightforward task to apply lemma \ref{lem:4} in the resonance region similarly to how we did it in the
non-resonance region.  Indeed, formulas \eqref{eq:41}, \eqref{eq:41n}, and \eqref{eq:43} are immediate corollaries of $|\bxi|^2=r(\bxi)^2+\Phi(\bxi)^2$, and \eqref{eq:45} follows from the standard results of perturbation theory (see, e.g., \cite{Kat})
applied to the operator pencil
$rA(\Phi)+B(\Phi)= r(A(\Phi)+r^{-1}B(\Phi))$.
The problem with this approach is that the coefficients $\hat a_j(\Phi)$ in
\eqref{eq:45} are not bounded in general. This unboundedness of the coefficients is caused by the fact that the eigenvalues of
$\hat B$ can be located very close to each other. However, lemma \ref{lem:tildeB} shows that the multiplicity
of any cluster of eigenvalues of $\hat B$ cannot be greater than $2$. This observation will be of a great help to us.

It will be slightly more convenient to introduce new operators $\BA=\BA_n:=R_n^2 A$ and
$\BB=\BB_n:=R_n^2 B$ (and $\tilde\BB:=R_n^2 \tilde B=\tilde P\BB\tilde P$); we will be assuming from now on that $R_n\leq\rho_n^{1/25}$. The reason for this change is that lemmas
\ref{lem:A} and \ref{lem:tildeB} can be reformulated in a more uniform way:
\bel\label{lem:BA}
There is a positive constant $C_2$ which satisfies two properties:
if $\mu$ is a non-zero eigenvalue of $\BA$, then $|\mu|\ge C_2$ and
\bee\label{eq:tildeBB}
\mu_{j+2}(\tilde \BB)-\mu_j(\tilde \BB)\ge 3C_2
\ene
uniformly over $j$, $n$, $l$ and $\bxi$.
\enl
\ber
1) It will be convenient to assume that $C_2<1/10$, which we will be doing from now on.

2) Of course, we have slightly better estimate for eigenvalues of $\mu_j(\tilde \BB)$. The distance between $\mu_{j+2}(\tilde \BB)$ and $\mu_j(\tilde \BB)$ is $\gg R_n^2$. But \eqref{eq:tildeBB} is enough for our purposes.
\enr
The importance of lemma \ref{lem:BA} can be seen from the following remark. Suppose that we could establish the
inequality \eqref{eq:tildeBB} with $\mu_{j+1}(\tilde \BB)$ instead of $\mu_{j+2}(\tilde \BB)$. Then, using the approach from the previous section,
we could prove that the coefficients $\hat a_j(\Phi)$ in \eqref{eq:45} are bounded, and this would
finish the proof of our main theorem. However, in general it could happen that two eigenvalues of $\tilde \BB$ lie close to each other.
Our further course of action will reflect this possibility. We will divide the segment $\CS=[-a,a]$
into two disjoint parts, $\CS=\tilde\CS\cup\hat\CS$. Roughly speaking, $\tilde\CS$ will be the region where the eigenvalues of
$\tilde\BB$ are far from each other, and $\hat\CS$ will be the region corresponding to couples of eigenvalues of
$\tilde\BB$ lying close to each other. To be more precise, we need yet more notation.
Let $\bxi\in\Xi_5(\bth)$ and $\boldeta=(\eta_1,\eta_2)\in\BPsi_0(\bxi)$  (recall that $\BPsi_0=\{\bxi+\Z\bth\}\cap\BUps(\bxi)$, so
$\tilde P=\CP^{(\bk)}(\BPsi_0)$; this means, in particular, that $\eta_1=\xi_1$). Then $(\eta_2)^2$ is an eigenvalue of the unperturbed operator $\tilde\BB_0(\bxi)$, say $(\eta_2)^2=\mu_{\tau(\boldeta)}(\tilde\BB_0)$.
Here, as above, we use the convention that if two eigenvalues $(\eta_2)^2$ and say $(\nu_2)^2$ coincide, we label them
according to the crystallographic order of their universal coordinates 
(of course, this could happen only if the
quasi-momentum $\hat k_2$ is either $0$ or $1/2$). Thus, we have defined a mapping $\tau:\BPsi_0(\bxi)\to\N$. Notice that
we can talk simply about the value $\tau(\boldeta)$, without specifying what $\bxi$ is, since if $\boldeta\in\BPsi_0(\bxi_j)$,
$j=1,2$, then $\BPsi_0(\bxi_1)=\BPsi_0(\bxi_2)$.
Next, for any point $\boldeta\in\BPsi_0(\bxi)$, we define
\bee\label{mappingh}
h(\boldeta):=\mu_{\tau(\boldeta)}(\tilde B(\bxi)).
\ene
and
\bee\label{mappingbh}
\bh(\boldeta):=\mu_{\tau(\boldeta)}(\tilde\BB(\bxi))=R_n^2h(\boldeta).
\ene
Then, we can reformulate \eqref{eq:tildeBB} like this: for each $\boldeta\in\BPsi_0(\bxi)$, there is at most one point
$\bnu\in\BPsi_0(\bxi)$, $\bnu\ne\boldeta$ such that $|\bh(\bnu)-\bh(\boldeta)|< 3C_2$. Notice that this whole construction
does not depend on the first coordinate $\eta_1$
(we can recall the paragraph after the proof of lemma \ref{lem:tildeB} at this stage), so we can think of $\tau$ as a mapping
$\tau:\eta_2\mapsto \tau(\boldeta)$, where $\boldeta$ is any point with second coordinate $\eta_2$ such that
$\boldeta\in\BPsi_0(\bxi)$ for some $\bxi\in\Xi_5(\bth)$. Then the domain of thus defined mapping $\tau$ is some interval $\tilde I$ which consist of all second coordinates $\eta_2$ of points
$\boldeta\in\BPsi_0(\bxi)$ with $\bxi\in\Xi_5(\bth)$; obviously,
$\tilde I\supset [-a,a]$. Similarly, $h:\eta_2\mapsto \mu_{\tau(\eta_2)}(\tilde\BB(\bxi))$ is a well-defined function on $\tilde I$.
Let $s$ be a small parameter which we will fix later on. At the moment, we put $s=\frac12 C_2$, but we will decrease $s$
later.
We define $\tilde I_1=\tilde I_1(s)$ to consist of all points $\eta_2$ from $\tilde I$ such that there exists a non-zero integer $m$
such that $\eta_2+m|\bth|\in\tilde I$ and
\bee\label{tildeI1}
\bigm|\,\bh(\eta_2)-\bh(\eta_2+m|\bth|)\,\bigm|\le s.
\ene
In other words, $\tilde I_1$ consists
of all points $\eta_2$ such that the eigenvalue of $\tilde\BB$ corresponding to $\eta_2$ is close to being multiple.
We also put $\tilde I_0=\tilde I_0(s):=\tilde I\setminus\tilde I_1(s)$. Let us study the properties of this partition. First of all, due
to lemma \ref{lem:BA}, for each $\eta_2\in\tilde I_1$, equation \eqref{tildeI1} is satisfied for precisely one value of $m$;
obviously, then $\eta_2+m|\bth|$ also belongs to $\tilde I_1$. Thus, we can define a mapping $\iota:\tilde I_1\to\tilde I_1$
by the formula $\iota(\eta_2)=\eta_2+m|\bth|$, where $m\ne 0$ is chosen so that \eqref{tildeI1} is satisfied. Obviously, then
$\iota^2=Id$. We can extend the mapping $\iota$ to the whole $\tilde I$ by requesting that $\iota(\eta_2)=\eta_2$ whenever
$\eta_2\in\tilde I_0$. Sometimes we will slightly abuse this notation by writing $\iota(\boldeta):=(\eta_1,\iota(\eta_2))$.
\bel
Suppose $\eta_2\in[-a,a]$. Then $\iota(\eta_2)\in[-a,a]$.
\enl
\bep
If $\eta_2\in\tilde I_0$, the statement is obvious. Suppose, $\eta_2\in(\tilde I_1\cap [-a,a])$. Without loss of generality we can assume that $\eta_2$ is positive. Notice that
$|h(\eta_2)-|\eta_2|^2|\le v$.
This implies that whenever $\eta_2\le a/2$, the statement holds. Suppose,
$\eta_2\ge a/2\gg\rho_n^{1/3}$. Then if \eqref{tildeI1} is satisfied, we have
\bee\label{tildeI2}
\bigm|\,|\eta_2|^2-|\eta_2+m|\bth||^2\,\bigm|\le s+2v,
\ene
and thus
\bee\label{tildeI3}
\bigm|\,|\eta_2|-|\eta_2+m|\bth||\,\bigm|\ll\rho_n^{-1/3}.
\ene
Since $\eta_2$ is assumed to be positive and $\eta_2+m|\bth|$ is negative (otherwise there is no chance for
\eqref{tildeI3} to hold), this means
\bee\label{tildeI4}
\bigm|\,\eta_2+\frac{m}{2}|\bth|\,\bigm|\ll\rho_n^{-1/3}.
\ene
Obviously, we will have the same inequality for $\iota(\eta_2)$:
\bee\label{tildeI5}
\bigm|\,|\iota(\eta_2)|+\frac{m}{2}|\bth|\,\bigm|\ll\rho_n^{-1/3},
\ene
so both $\eta_2$ and $\iota(\eta_2)$ are close to (i.e. within distance $o(1)$) points of the form $\pm \frac{m}{2}|\bth|$.
Now the second condition \eqref{conditionsa} implies that the distance from $a$ to any point of the form
$\pm\frac{m}{2}|\bth|$ is at least $|\bth|/4$. Thus, if $|\eta_2|<a$, this implies that $|\iota(\eta_2)|<a$.
\enp
Now let us establish the relationship between the labeling $\tau:\BPsi_0(\bxi)\to\N$ used to define mapping $h$ and the
labeling $t:\BUps(\bxi)\to\N$ defined by \eqref{mappingt}.
\bel
Let $\tilde T$ be the number of elements in $\BUps(\bxi)$ whose first coordinate is strictly less than $\xi_1$. Then for each
$\boldeta\in\BPsi_0(\bxi)$ we have
\bee\label{mappingttau}
t(\boldeta)=\tau(\boldeta)+\tilde T.
\ene
\enl
\bep
It follows from the proof of lemma \ref{lem:Xi5} that whenever $\bmu\in \BUps(\bxi)\setminus\BPsi_0(\bxi)$ and
$\boldeta\in\BPsi_0(\bxi)$, the following
two conditions are equivalent: $\mu_1<\eta_1$ and $|\bmu|<|\boldeta|$. Indeed, to say that $\bmu\not\in\BPsi_0(\bxi)$
is equivalent to saying that $\mu_1\ne\eta_1$. Suppose that $\mu_1<\eta_1$. then $\eta_1-\mu_1\gg R_n^{-2}$, so
$(\eta_1)^2-(\mu_1)^2\gg \rho_n R_n^{-2}\gg\rho_n^{4/5}$. Since $|\mu_2|\ll a\ll \rho_n^{1/3}$, this implies
$|\bmu|<|\boldeta|$. The case $\mu_1>\eta_1$ is treated similarly.
The rest follows from the definitions of
mappings $t$ and $\tau$.
\enp

Now let us recall that because of \eqref{newpencil}, we are interested in studying eigenvalues of the operator
pencil
\bee\label{BZ}
\BZ(r):=r\BA+\BB=r(\BA+r^{-1}\BB)=R_n^2Z(r),
\ene
where
\bee\label{Z}
Z(r)=rA+B,
\ene
$r=\xi_1$, and operators $\BA$ and $\BB$ depend on $\Phi(\bxi)=\xi_2$ for some point $\bxi=(\xi_1,\xi_2)\in\Xi_5(\bth)$.
These operators act in the Hilbert space which we have denoted by $\GH(\Phi(\bxi))$; see
the paragraph after the proof of lemma \ref{lem:tildeB} 
for the discussion
of this Hilbert space. To be more precise, we fix a point $\boldeta\in\BUps(\bxi)$ and study the function $g(\boldeta)$.
We are only interested in eigenvalues of $Z(r)$ which are bounded as $r\to\infty$ (which means, they can be considered as
perturbations of zero eigenvalues of $A$ or, equivalently, that $\boldeta\in\BPsi_0(\bxi)$). The first order of approximation of such eigenvalues as $r\to\infty$ are
eigenvalues of $\tilde\BB$.
We will consider separately two cases: $\eta_2\in\tilde I_1$ and $\eta_2\in\tilde I_0$. The former case
is much more difficult, and we give all necessary details. The latter case is much simpler and can be treated
analogously to the first case (alternatively, one can apply methods similar to those we used in the previous section);
we will make some remarks on this case later. So, let us assume that $\eta_2\in\tilde I_1$.

Recall that $\tilde P$ is a projection
onto $\ker A=\ker\BA$; we also denote $P':=P(\bxi)-\tilde P$.
Let $P_0\,(<\tilde P)$ be projector onto span of
two eigenfunctions of $\tilde\BB$ corresponding to $\bh(\eta_2)$ and $\bh(\iota(\eta_2))$; put $P_0':=\tilde P-P_0$.
By $\mu$ we denote a spectral parameter, which at the moment we assume satisfies $|\mu|\leq 2\|\BB\|$.
Operator $P'\BA P'+\tilde P$ is invertible and $\|(P'\BA P'+\tilde P)^{-1}\|\leq c$ with constant $c>0$ uniform
with respect to $n$ and $\eta_2$. Thus, for sufficiently large $\rho_0$, operator
\begin{equation}
S_\mu:=P'\BA P'+\tilde P+\frac1r(\BB-\mu)
\end{equation}
is invertible. We have
\begin{equation}
r\BA+\BB-\mu=(I-\tilde PS_\mu^{-1})rS_\mu.
\end{equation}
Then $\mu$ is an eigenvalue of $r\BA+\BB$ if and only if $0$ is an
eigenvalue of $I-\tilde PS_\mu^{-1}$. Obviously, corresponding eigenfunction $z$ belongs
to $\tilde P \GH(\Phi)$. Thus,
\begin{equation}\label{one}
z=S_\mu^{-1}z+y
\end{equation}
for some $y\in P'\GH(\Phi)$. We have $S_\mu z-z=S_\mu y$. Then
\begin{equation}\label{two}
P'S_\mu \tilde Pz=P'S_\mu P'y,\ \ \ y=(P'S_\mu P')^{-1}P'S_\mu \tilde Pz.
\end{equation}
Thus,
\begin{equation}
\begin{split}\label{three}
& 0=(\tilde PS_\mu \tilde P-\tilde P-\tilde PS_\mu P'(P'S_\mu P')^{-1}P'S_\mu \tilde P)z=\cr &
\frac1r(\tilde P(\BB-\mu)\tilde P-\frac1r\tilde P(\BB-\mu)P'(P'\BA P'+\frac1rP'(\BB-\mu)P')^{-1}P'(\BB-\mu)\tilde P)z.
\end{split}
\end{equation}
Therefore, zero is an eigenvalue of operator $\tilde P(\BB-\mu)\tilde P-\frac1r K_\mu$, where
\begin{equation}\label{four}
K_\mu=\tilde P(\BB-\mu)P'(P'\BA P'+\frac1r P'(\BB-\mu)P')^{-1}P'(\BB-\mu)\tilde P.
\end{equation}
Note that since $\tilde P\BB P'=R_n^2 \tilde PVP'$ we have
\begin{equation}
\|K_\mu\|+\sum\limits_{j=1}^N\|\frac{d^j}{d\mu^j}K_\mu\|\leq C(N)R_n^4,
\end{equation}
where $C(N)$ depends only on $N$ and $V$, provided $r\geq\rho_n^{3/4}$ and $\rho_0(V)$ is sufficiently large.


Next, we want to narrow the range of $\mu$'s which serve as the candidates for being the eigenvalues of $r\BA+\BB$.
Let us at the moment only look for eigenvalues $\mu$ such that
\bee\label{mu}
|\mu-\bh(\eta_2)|\leq C_2.
\ene
If we assume that \eqref{mu} is satisfied,
the operator $\tilde P(\BB-\mu)\tilde P + P_0$
is invertible (on $\GV$) and inverse operator is bounded uniformly in $n$ and $\eta_2$.
We have
\begin{equation}
\tilde P(\BB-\mu)\tilde P-\frac1r K_\mu=(I-P_0D_\mu^{-1})D_\mu,
\end{equation}
where $D_\mu:=\tilde P(\BB-\mu)\tilde P+P_0-\frac1r K_\mu$.

Now we repeat the same construction as in \eqref{one}--\eqref{four}, only with respect to the pair of projections $P_0,\tilde P$ instead of $\tilde P, P(\bxi)$. As a result,
we obtain that $\mu$ is an eigenvalue of $\BZ(r)=r\BA+\BB$ if and only if
zero is an eigenvalue of the operator $P_0(\BB-\mu)P_0-G_\mu$, where
\begin{equation}
G_\mu=\frac1r (P_0K_\mu P_0+\frac1r P_0K_\mu P_0'(P_0'(\BB-\mu)P_0'-\frac1r P_0'K_\mu P_0')^{-1}P_0'K_\mu
P_0).
\end{equation}
The operator $P_0(\BB-\mu)P_0-G_\mu$ is, in fact, a $(2\times2)$-matrix. Note that in a suitable basis, $P_0(\BB-\mu)P_0$ is a diagonal matrix with $\bh(\eta_2)-\mu$ and $\bh(\iota(\eta_2))-\mu$ standing on the diagonal.
Calculating the determinant of $P_0(\BB-\mu)P_0-G_\mu$ in this basis, we obtain that $\mu$ satisfying \eqref{mu} is an
eigenvalue of $\BZ(r)$ if and only if
\begin{equation}\label{determinant}
(\bh(\eta_2)-\mu+\frac1r \al_1)(\bh(\iota(\eta_2))-\mu+\frac1r \al_2)-\frac{1}{r^2}\beta^2=0.
\end{equation}
Here, $\al_1$, $\al_2$, and $\beta$ are functions of $\mu$ and $r$ (depending on $\eta_2$ as a parameter) analytic in
$|\mu-\bh(\eta_2)|<C_2,\ r>\rho_n^{3/4}$ and satisfying
\begin{equation}\label{estimate7}
|\al_1|+|\al_2|+|\beta|+\sum\limits_{j=1}^N\left(|\frac{d^j\al_1}{d\mu^j}|+|\frac{d^j\al_2}{d\mu^j}|+|\frac{d^j\beta}{d\mu^j}|\right)\leq C(N)R_n^4,
\end{equation}
with constant $C(N)$ uniform in $n$ and $\eta_2\in{\tilde I}_1$, provided
$\rho_0$ is sufficiently large. We put $\nu:=\mu-\bh(\eta_2)$, $\epsilon:=\frac{\rho_n^{3/4}}{r}$, and $\de:=\bh(\eta_2)-\bh(\iota(\eta_2))$.  Then \eqref{determinant} is equivalent to
\begin{equation}\label{eq:roma}
F(\nu,\delta,\epsilon;\eta_2
):=\nu^2+\nu(\delta-\epsilon\frac{\al_1+\al_2}{\rho_n^{3/4}})-\epsilon\delta
\frac{\al_1}{\rho_n^{3/4}}+\epsilon^2\frac{\al_1\al_2-\beta^2}{\rho_n^{3/2}}=0.
\end{equation}
Here, we temporarily consider $\delta$ as independent variable, and $\al_1$, $\al_2$, and $\beta$ are considered as functions of $\nu$
and $\epsilon$, depending on $\eta_2$ and $\rho_n$ as parameters.
It follows
from corollary \ref{cor:WPT1} that there exists a neighborhood $\omega$ of $(\nu,\delta,\epsilon)=(0,0,0)$ such that $F=0$
in $\omega$ if and only if
\begin{equation}\label{eq:roma1}
\nu^2-2X_1(\delta,\epsilon;\eta_2)\nu+X_2(\delta,\epsilon;\eta_2)=0.
\end{equation}
Here, $X_1,\ X_2$ are analytic in $\delta,\ \epsilon$ in $\omega$ and
$X_1(0,0;\eta_2)=X_2(0,0;\eta_2)=0$ (the difference between \eqref{eq:roma1} and \eqref{eq:roma} is that functions $X_1$, $X_2$ do not depend on $\nu$). Moreover, it follows from
corollary \ref{cor:WPT2} and uniformness of our estimates that $\omega$ can be
chosen to depend on $V$ only; we also can achieve that $\omega$ contains the set $|\epsilon|<\epsilon_0,\,|\delta|<\delta_0$,
where, $\epsilon_0$ and $\delta_0$ do not depend on  $n$ and $\eta_2\in{\tilde I}_1$ (they depend only on $V$).
We also have uniform upper bounds for $X_1,\ X_2$
and its derivatives. Indeed, we have uniform upper bound \eqref{bound}. Then analyticity of $W$ (which is equal to
$\nu^2-2X_1(\delta,\epsilon;\eta_2)\nu+X_2(\delta,\epsilon;\eta_2)$ in our case), \eqref{bound}, and Cauchy's integral formula imply upper bounds for the coefficients $X_1$, $X_2$.
Now we can solve
quadratic equation \eqref{eq:roma1} and obtain $\nu_{1,2}=X_1\pm\sqrt{X_1^2-X_2}$, or
$\mu_{1,2}=\bh(\eta_2)+X_1\pm\sqrt{X_1^2-X_2}$. Thus, recalling that $H(\bxi)=r^2+Z(r)=r^2+R_n^{-2}\BZ(r)$, we deduce that the points
\bees
r^2+R_n^{-2}\bh(\eta_2)+R_n^{-2}(X_1\pm\sqrt{X_1^2-X_2})=r^2+h(\eta_2)+R_n^{-2}(X_1\pm\sqrt{X_1^2-X_2})
\enes
are eigenvalues of $H(\bxi)$; notice that, since $H(\bxi)$ is a self-adjoint operator, this implies that
$X_1^2-X_2\geq0$ when all the variables take real values. Now, the definition of the mapping $g$ implies
that there exist two points, $\bal_+,\bal_-\in\BUps(\bxi)$ such that $g(\bal_\pm)=r^2+h(\eta_2)+R_n^{-2}(X_1\pm\sqrt{X_1^2-X_2})$.
Lemma \ref{lem:Xi5} implies that $\bal_\pm\in\BPsi_0(\bxi)$.

Now we recall that $\delta$ in fact is not an independent parameter, but $\delta=\bh(\eta_2)-\bh(\iota(\eta_2))$.
Then our functions $X_1,\ X_2$ will be analytic in $\epsilon$ for
$|\epsilon|<\epsilon_0$, provided
$|\bh(\eta_2)-\bh(\iota(\eta_2))|<\delta_0$. Thus, we have proved
the following statement: \bel\label{gh} There exist positive numbers
$\rho_0$, $\delta_0$ and $\epsilon_0$  and two functions $X_1$,
$X_2$, $X_j=X_j(\delta,\epsilon;\eta_2)$, such that $X_j$ are
analytic in $\delta$ and $\epsilon$ when $|\delta|<\delta_0$ and
$|\epsilon|<\epsilon_0$ and the following property is satisfied.
Suppose, $\bxi=(\xi_1,\xi_2)\in\Xi_5(\bth)$ and
$\boldeta\in\BPsi_0(\bxi)$ with $\eta_2\in\tilde I_1(s)$,
$s<\min(\delta_0,\frac12 C_2)$. Then, there exist two points
$\bal_{\pm}=\bal_\pm(\eta_2)\in\BPsi_0(\bxi)$ such that
\bee\label{eq:gh}
g(\bal_\pm)=r^2+h(\eta_2)+R_n^{-2}(X_1\pm\sqrt{X_1^2-X_2}), \ene
where
$X_j=X_j(h(\eta_2)-h(\iota(\eta_2)),\frac{\rho_n^{3/4}}{r};\eta_2)$.
Moreover, each point $\bnu\in\BPsi_0(\bxi)\cap\tilde I_1$ can be
expressed as $\bnu=\bal_\pm(\mu_2)$ for some $\bmu\in\BUps_0(\bxi)$.
\enl \bep The last statement is the only one we have not proved so
far. However, it follows from the standard pigeonhole arguments
based on the fact that the number of pairs $(\bal_+,\bal_-)$ is the
same as the number of pairs $(\eta_2,\iota(\eta_2))$ with
$\eta_2\in\tilde I_1(s)$. \enp \ber Instead of assuming that $n$
(and thus $\rho_n$) is sufficiently large, we prefer to assume that
$\rho_0$ is large enough and prove that estimates hold uniformly for
all $n$ (recall that $\rho_n=2^n\rho_0$). Also, since we always have
$\delta=\bh(\eta_2)-\bh(\iota(\eta_2))$ depending only on $\eta_2$,
we will often skip mentioning the dependence of the functions $X_j$
on the first variable and write $X_j=X_j(\epsilon;\eta_2)$. \enr
Note that $X_j$ and their derivatives enjoy uniform upper bounds and
we also have $X_j(0,0;\eta_2)=0$. Therefore, by decreasing the
values of $\delta_0$ and $\epsilon_0$ if necessary, we can achieve
that $|X_1\pm\sqrt{X_1^2-X_2}|<C_2/10$ when $|\delta|<\delta_0$ and
$|\epsilon|<\epsilon_0$ uniformly over $\eta_2$. Now we can fix the
value of the parameter $s$ which we used to define the sets $\tilde
I_1(s)$ and $\tilde I_0(s)$: we put
$s:=\min(\frac12\delta_0,\frac14C_2)$.


Suppose now that $\boldeta\in\tilde I_0(s)$. We have:
\bel\label{gh1}
There exist positive numbers $\rho_0$ and $\epsilon_0$  and a function $Y=Y(\epsilon;\eta_2)$, such that
$Y$ is analytic in $\epsilon$ when 
$|\epsilon|<\epsilon_0$ and the following property is satisfied.
Suppose, $\bxi=(\xi_1,\xi_2)\in\Xi_5(\bth)$ and $\boldeta\in\BPsi_0(\bxi)$ with $\eta_2\in\tilde I_0(s)$. Then, there
exists a point $\bbeta=\bbeta(\eta_2)\in\BPsi_0(\bxi)$ such that $g(\bbeta)=r^2+h(\eta_2)+R_n^{-2}Y(\frac{\rho_n^{3/4}}{r};\eta_2)$.
\enl
The proof is similar to the above and is even simpler; in fact, this proof essentially is equivalent to the
proof of lemma 6.1 from \cite{Par}. That is why we
just give the sketch of the proof and make some remarks on uniformness.
We start with the formula \eqref{four}. Now, $P_0$ is a projector onto
{\it one-dimensional} subspace corresponding to eigenvalue $\bh(\eta_2)$
of $\tilde\BB$. We consider $\mu$ such that (cf. \eqref{mu})
$$|\mu - \bh(\eta_2)|\leq s/3.$$
Then operator ${\tilde P}(\BB - \mu){\tilde P}+P_0$ is invertible. If necessary we increase $\rho_0$ to ensure
that operator $D_\mu$ is invertible. Next, we repeat all further arguments from the proof of lemma \ref{gh}
which are simpler in this case since $G_\mu$ is a scalar-valued function now. We obtain that $\mu$
is an eigenvalue of $\BZ(r)$ if and only if
$$
{\tilde F}:=\bh(\eta_2)-\mu+\frac1r\alpha=0,
$$
where $\alpha$ is analytic in $|\mu - \bh(\eta_2)|\leq s/3$, $r>\rho_n^{3/4}$
and depends on $\eta_2$ as a parameter. It also satisfies estimate
similar to \eqref{estimate7} uniformly in $n$ and $\eta_2\in{\tilde
I}_0$. Applying Corollary~\ref{cor:WPT1} (alternatively, we can just use the implicit function theorem)
we obtain that ${\tilde F}=0$ in some neighborhood $\omega$ of $(\mu,\frac{\rho_n^{3/4}}{r})=(\bh(\eta_2),0)$
if and only if $\mu=\bh(\eta_2)+Y(\frac{\rho_n^{3/4}}{r};\eta_2)$ for some analytic function $Y$. The neighbourhood $\omega$ contains the set
$|\frac{\rho_n^{3/4}}{r}|=|\epsilon|<\epsilon_0$, where $\epsilon_0$ does not depend on $n$ and $\eta_2\in{\tilde I}_0$.

As above, we have $Y(0;\eta_2)=0$ and $Y$ and its
derivatives being uniformly bounded (actually, the bound depends on $\delta_0$ only). Thus, by decreasing
$\epsilon_0$ again if necessary, we can achieve that $|Y|<s/3$ whenever $|\epsilon|<\epsilon_0$.

\bel

(A) Suppose, $\eta_2\in\tilde I_1(s)$. Then we either have $\bal_1=\boldeta$ and $\bal_2=\iota(\boldeta)$, or
$\bal_2=\boldeta$ and $\bal_1=\iota(\boldeta)$.

(B) Suppose, $\eta_2\in\tilde I_0(s)$. Then we have $\bbeta=\boldeta$.
\enl
\bep

Suppose, $\boldeta$ and $\bnu$ are two different points from $\BPsi_0(\bxi)$ and $\bnu\ne\iota(\boldeta)$.
Suppose for definiteness that $\tau(\boldeta)<\tau(\bnu)$, i.e. that $h(\eta_2)<h(\nu_2)$. Then:

(a) if both $\eta_2$ and $\nu_2$ belong to $\tilde I_0$, we have $\bh(\nu_2)-\bh(\eta_2)>s$, so
\bee
\bh(\eta_2)+Y(\cdot;\eta_2)<\bh(\nu_2)+Y(\cdot;\nu_2),
\ene
and thus $g(\bbeta(\boldeta))<g(\bbeta(\bnu))$.

(b) if both $\eta_2$ and $\nu_2$ belong to $\tilde I_1$, we have $\bh(\nu_2)-\bh(\eta_2)\geq3C_2$, so
\bee
\bh(\eta_2)+X_1(\cdot;\eta_2)\pm\sqrt{X_1^2(\cdot;\eta_2)-X_2(\cdot;\eta_2)}< \bh(\nu_2)
+X_1(\cdot;\nu_2)\pm\sqrt{X_1^2(\cdot;\nu_2)-X_2(\cdot;\nu_2)},
\ene
and thus $g(\bal_±(\boldeta))<g(\bal_±(\bnu))$.

(c) finally, if we have say $\eta_2\in\tilde I_1(s)$ and $\nu_2\in\tilde I_0(s)$, we have $\bh(\nu_2)-\bh(\eta_2)\geq3C_2$, so
\bee
\bh(\eta_2)+X_1(\cdot;\eta_2)\pm\sqrt{X_1^2(\cdot;\eta_2)-X_2(\cdot;\eta_2)}< \bh(\nu_2)
+Y(\cdot;\nu_2),
\ene
and thus $g(\bal_±(\boldeta))<g(\bbeta(\bnu))$.

In all these cases, we have $t(\boldeta)<t(\bnu)$. Now the proof follows from the pigeonhole argument.
\enp



\bec\label{cor:7.12}
Let $\bxi=(\xi_1,\xi_2)\in\Xi_5(\bth)$.

(a) Suppose, $\boldeta\in\BPsi_0(\bxi)$ with $\eta_2\in\tilde I_1(s)$. Then,
we either have
\bee\label{eq:gh2}
g(\boldeta)=r^2+h(\eta_2)+R_n^{-2}(X_1+\sqrt{X_1^2-X_2})
\ene
and
\bee\label{eq:gh3}
g(\iota(\boldeta))=r^2+h(\eta_2)+R_n^{-2}(X_1-\sqrt{X_1^2-X_2}),
\ene
or
\bee\label{eq:gh4}
g(\boldeta)=r^2+h(\eta_2)+R_n^{-2}(X_1-\sqrt{X_1^2-X_2})
\ene
and
\bee\label{eq:gh5}
g(\iota(\boldeta))=r^2+h(\eta_2)+R_n^{-2}(X_1+\sqrt{X_1^2-X_2}).
\ene

(b) Suppose,  $\boldeta\in\BPsi_0(\bxi)$ with $\eta_2\in\tilde I_0(s)$. Then, we have
\bee\label{eq:gh6}
g(\boldeta)=r^2+h(\eta_2)+R_n^{-2}Y(\frac{\rho_n^{3/4}}{r};\eta_2).
\ene
\enc
\ber
Suppose, $\boldeta$ and $\boldeta'$ are two different points with the same second coordinate $\eta_2\in\tilde I_1$.
Then we either have both $t(\boldeta)>t(\iota(\boldeta))$ and $t(\boldeta')>t(\iota(\boldeta'))$, or
both $t(\boldeta)<t(\iota(\boldeta))$ and $t(\boldeta')<t(\iota(\boldeta'))$. This shows that we either
have \eqref{eq:gh2}--\eqref{eq:gh3} or \eqref{eq:gh4}--\eqref{eq:gh5} simultaneously for both
$\boldeta$ and $\boldeta'$.
\enr

Since the derivative of $Y$ is bounded in
$\{|\epsilon|<\epsilon_0\}$, expression \eqref{eq:gh6} is increasing
function of $r$ (assuming, as we always do, that $\rho_0$ is
sufficiently large). We denote by $q=q(\eta_2)\,(=q(\Phi))$ the
value of $r$ which makes the RHS of the equation \eqref{eq:gh6}
equal to $\rho^2$. Unfortunately, the same argument will not work
with expressions \eqref{eq:gh2} or \eqref{eq:gh4} (when we
differentiate the RHS of these formulas, we obtain square root in
the denominator). It turns out, however, that if we fix $\eta_2$,
the equations \bee\label{verynew1}
r^2+h(\eta_2)+R_n^{-2}(X_1+\sqrt{X_1^2-X_2})=\rho^2 \ene and
\bee\label{verynew2}
r^2+h(\eta_2)+R_n^{-2}(X_1-\sqrt{X_1^2-X_2})=\rho^2 \ene have
exactly one solution each. Indeed, the intermediate value theorem
implies that there is at least one solution to each equation, and
later on in remark \ref{verynewrem} we will see that the total
number of solutions of \eqref{verynew1} and \eqref{verynew2} is at
most two. We denote by $q=q(\eta_2)\,(=q(\Phi))$ the value of $r$
which makes the RHS of the relevant equation \eqref{eq:gh2} or
\eqref{eq:gh4} equal to $\rho^2$.

Then similarly to our proof of lemma \ref{lem:4} (more precisely, of equation \eqref{eq:416}), we obtain the following formula:
\bee\label{eq:44n}
\vol(\hat A^+\cap\CA^{(n)}_{L+l})-\vol(\hat A^-\cap\CA^{(n)}_{L+l})=\int^a_{-a}(q(\eta_2)-\sqrt{\rho^2-\eta_2^2})d\eta_2.
\ene

Thus, in order to compute $\vol(\hat A^+\cap\CA^{(n)}_{L+l})-\vol(\hat A^-\cap\CA^{(n)}_{L+l})$, we
need to compute $q(\eta_2)$.
We will consider the case where $\eta_2\in \tilde I_1$
(another case is simpler and can be dealt with in the same way). We also assume for definiteness that
formulas \eqref{eq:gh2}--\eqref{eq:gh3} are the valid ones, so
we need to solve equation
\bee\label{eq:inverse1}
r^{2}+
h(\eta_2)+R_n^{-2}(X_1+\sqrt{X_1^2-X_2})
=\la.
\end{equation}
Thus, $q(\eta_2)$ is the (only) value of $r$ which makes \eqref{eq:inverse1} valid, and
$q(\iota(\eta_2))$ is the (only) value of $r$ which solves the following equation:
\bee\label{eq:inverse2}
r^{2}+
h(\eta_2)+R_n^{-2}(X_1-\sqrt{X_1^2-X_2})
=\la.
\end{equation}
Now we introduce a new unknown variable
\bee
\sigma:=\frac{r}{\rho}-1,
\ene
so that
\begin{equation}
r=\rho(1+\sigma);
\end{equation}
we also put
\bee
{\tilde \epsilon}:=\rho_n^{3/4}/\rho.
\ene
Then direct calculations show that \eqref{eq:inverse1} is equivalent to
\begin{equation}\label{kvadrat}
\begin{split}
& (\sigma^2+2\sigma)+\rho_n^{-3/2}{\tilde \epsilon}^{2}(h(\eta_2)+R_n^{-2}X_1({\tilde
\epsilon}(1+\sigma)^{-1};\eta_2))=\cr
& -\rho_n^{-3/2}R_n^{-2}{\tilde \epsilon}^{2}\sqrt{X_1^2({\tilde
\epsilon}(1+\sigma)^{-1};\eta_2)-X_2({\tilde \epsilon}(1+\sigma)^{-1};\eta_2)}.
\end{split}
\end{equation}
Taking square of the last equality we obtain
\begin{equation}
W(\sigma,{\tilde \epsilon};\eta_2):=(\sigma+2)^2 \sigma^2+{\tilde \epsilon}^{2}\rho_n^{-3/2}w(\sigma,{\tilde
\epsilon};\eta_2)=0,
\end{equation}
where $w$ is a certain function;
the properties of $w$ follow from the properties of $X_j$. In particular, $w$ is analytic in $|\sigma|<1/2$ and
$|{\tilde \epsilon}|<\epsilon_0 /2$. Moreover, the bounds for $w$ and
its derivatives are uniform in $n$ and $\eta_2\in {\tilde I}_1$. We see that
$$
W(0,0;\eta_2)=W'_\sigma(0,0;\eta_2)=0\ \  \hbox{and}\ \ W''_{\sigma\sigma}(0,0;\eta_2)=8\not=0.
$$
Applying again theorem \ref{WPT}, we obtain that (in the
neighborhood of $(\sigma,{\tilde \epsilon})=(0,0)$)
$W(\sigma,{\tilde \epsilon};\eta_2)=0$ if and only if
\bee\label{eq:sigma1} \sigma^2-2X_3({\tilde
\epsilon};\eta_2)\sigma+X_4({\tilde \epsilon};\eta_2)=0, \ene where
$X_j({\tilde \epsilon};\eta_2)$ are analytic in ${\tilde \epsilon}$
and $X_j(0;\eta_2)=0$ for $j=3,4$. \ber\label{verynewrem} Since the
equation \eqref{eq:sigma1} has two $\sigma$-solutions, this implies
that the total number of solutions of \eqref{verynew1} and
\eqref{verynew2} is at most two. \enr

The solutions of \eqref{eq:sigma1} are $\sigma_1({\tilde \epsilon};\eta_2):=X_3+\sqrt{X_3^2-X_4}$ and
$\sigma_2({\tilde \epsilon};\eta_2):=X_3-\sqrt{X_3^2-X_4}$. Thus, we either have $q(\eta_2)=\rho(1+\sigma_1)$
and $q(\iota(\eta_2))=\rho(1+\sigma_2)$, or $q(\eta_2)=\rho(1+\sigma_2)$
and $q(\iota(\eta_2))=\rho(1+\sigma_1)$; for the sake of definiteness we assume the former possibility.

Put
$$
T(\eta_2,\rho)=T_n(\eta_2,\rho):=q(\eta_2)+q(\iota(\eta_2))={\rho}(2+\sigma_1+\sigma_2)=2\rho (1+X_3).
$$
According to Corollary~\ref{cor:WPT2}, $X_3$ and consequently $\rho^{-1}T_n$ is
analytic in ${\tilde \epsilon}$ for $|{\tilde \epsilon}|<c(V)$ with some constant $c(V)>0$ uniform in $n$ and
$\eta_2\in {\tilde I}_1$. Function $\rho^{-1}T_n$ and its derivatives are
bounded uniformly in $n$ and $\eta_2\in {\tilde I}_1$.

Thus, for $\eta_2\in {\tilde I}_1$ we
have $\rho^{-1}T_n$ is analytic (while $\sigma_1,\ \sigma_2$ are only algebraic).

Now assume that $\eta_2\in {\tilde I}_0$. As usual, this case is similar to the case $\eta_2\in {\tilde I}_1$, but
simpler. We
solve equation
\begin{equation}\label{kvadrat1}
r^2+h(\eta_2)+R_n^{-2}Y(\frac{\rho_n^{3/4}}{r};\eta_2)=\la,
\end{equation}
where $Y$ is analytic in $1/r$. Using arguments
similar to the first case, we obtain that \eqref{kvadrat1} has a unique solution $r=:q(\eta_2)$.
We define $T(\eta_2,\rho)=T_n(\eta_2,\rho):=2q(\eta_2)$ for $\eta_2\in {\tilde I}_0$.

Next, notice that $|{\tilde \epsilon}|<c(V)$ is satisfied if
$r>\rho_n^{3/4}2(c(V))^{-1}=\rho_n^{4/5}(\rho_n^{1/20}c(V)/2)^{-1}$. Thus if we assume that $\rho_0$ is large enough, we ensure that the set $\{r\ge\rho_n^{4/5}\}$ is included into domains of analyticity of all
our analytic functions.

According to \eqref{eq:44n} we have
\begin{equation}\label{dos}
\begin{split}
& \vol(\hat A^+\cap\CA^{(n)}_{L+l})-\vol(\hat A^-\cap\CA^{(n)}_{L+l})=\cr
& \int^a_{-a}(q(\eta_2)-\sqrt{\rho^2-\eta_2^2})d\eta_2=\frac12\int^a_{-a}(T(\eta_2)-2\sqrt{\rho^2-\eta_2^2})d\eta_2.
\end{split}
\end{equation}
Note that although $q(\eta_2)$ is not an analytic function of $\rho$ (it involves square root of analytic functions),
the function $T(\eta_2)$ is analytic.

The proof is almost finished, since the RHS of \eqref{dos} is analytic in $\rho$ for sufficiently large $\rho$ (recall that $|\eta_2|\ll a\ll\rho_n^{1/3}$). The only remaining thing is to obtain
some estimates for coefficients in the analytic expansion of $T_n$ (and thus of \eqref{dos}). We have
\begin{equation}
T_n=\rho\sum\limits_{j=0}^\infty t_{j}(n,\eta_2)\frac{1}{\rho^j}.
\end{equation}
It easily follows from \eqref{kvadrat}, \eqref{kvadrat1} and definition of $T_n$ that $t_0=2$ and $t_1=0$.
Since $\rho^{-1}|T_n|\leq C$ uniformly in $\eta_2$ for any $\rho\ge\rho_n^{4/5}$, we obtain
\begin{equation}
|t_{j}(n,\eta_2)|\leq C'\rho_n^{4j/5}
\end{equation}
with constant $C'>0$ uniform in $n$ and $\eta_2$. Substituting it into \eqref{dos}, we derive
\begin{equation}\label{dos1}
\vol(\hat A^+\cap\CA^{(n)}_{L+l})-\vol(\hat A^-\cap\CA^{(n)}_{L+l})=\frac12\rho\sum\limits_{j=2}^\infty {\tilde e}_{j}(n)\frac{1}{\rho^{j}},
\end{equation}
where ${\tilde e}_{j}(n)=\int\limits_{-a}^{a}(t_{j}(n,\eta_2)-{\tilde t}_j(\eta_2))d\eta_2$. Here, we denoted by ${\tilde t}_j$ coefficients in analytic expansion
$$
2\sqrt{\rho^2-\eta_2^2}=\rho\left(2+\sum\limits_{j=2}^\infty {\tilde t}_j(\eta_2)\frac1{\rho^j}\right).
$$

We have
\begin{equation}
|{\tilde e}_{j}|\leq 4C'\rho_n^{4j/5+1/3}.
\end{equation}
Next,
\begin{equation}
\rho_n\sum\limits_{j=6M}^{\infty}|{\tilde e}_{j}(n)|\frac{1}{\rho_n^{j}}\leq C\rho_n^{4/3}\sum\limits_{j=6M}^{\infty}\rho_n^{-j/5}\leq
C\rho_n^{-6M/5+4/3}\leq C\rho_n^{-M}
\end{equation}
with constant $C>0$ uniform in $n$. Put $e_j(n):=\frac12{\tilde e}_{j+1}(n)$. Thus, using \eqref{dos1} and
estimates of the coefficients obtained above, we arrive at
\bel\label{final}
\begin{equation}
\vol(\hat A^+\cap\CA^{(n)}_{L+l})-\vol(\hat A^-\cap\CA^{(n)}_{L+l})=\sum\limits_{j=1}^{6M} e_j(n)\rho^{-j}+O(\rho_n^{-M})\ \ \ \hbox{for}\ \rho\in[\rho_n,4\rho_n],
\end{equation}
with $|e_j(n)|\ll\rho_n^{4j/5+6/5}$.
\enl
Lemma \ref{main_lem} now follows after summation over $l$ from lemma \ref{final}, lemma \ref{nonresonance}, lemma \ref{lem:new1}
and corollary \ref{cor:3}.
This finishes the proof of lemma \ref{main_lem} and, therefore, of theorem \ref{main_thm}.
\bibliographystyle{amsplain}

\providecommand{\bysame} {\leavevmode\hbox
to3em{\hrulefill}\thinspace}

\end{document}